\documentclass{aastex}

\shorttitle{Optical Monitoring of 3C 273}
\shortauthors{Xiong et
al.}

\begin{document}
\title{MULTI-COLOR OPTICAL MONITORING OF THE QUASAR 3C 273 FROM 2005 TO 2016}

\author{Dingrong Xiong, Jinming Bai,}
\affil{Yunnan Observatories,
Chinese Academy of Sciences, 396 Yangfangwang, Guandu District, Kunming, 650216, P. R. China}
\affil{Center for Astronomical Mega-Science, Chinese Academy of Sciences, 20A Datun Road, Chaoyang District, Beijing, 100012, P. R. China}
\affil{Key Laboratory for the Structure and Evolution of Celestial
Objects, Chinese Academy of Sciences, 396 Yangfangwang, Guandu District, Kunming, 650216, P. R. China}\email{xiongdingrong@ynao.ac.cn}

\author{Haojing Zhang,}
\affil{Department of Physics, Yunnan Normal University, Kunming
650500, China}\email{kmzhanghj@163.com}

\author{Junhui Fan,}
\affil{Center for Astrophysics, Guangzhou University, Guangzhou 510006, China}
\affil{Astronomy Science and Technology Research Laboratory of Department of Education of Guangdong Province, Guangzhou 510006, China}

\author{Minfeng Gu,}
\affil{Key Laboratory for Research in Galaxies and Cosmology, Shanghai Astronomical Observatory, Chinese Academy of Sciences, 80 Nandan Road, Shanghai 200030, China}

\author{Tingfeng Yi and Xiong Zhang}
\affil{Department of Physics, Yunnan Normal University, Kunming
650500, China}

\begin{abstract}

We have monitored the quasar 3C 273 in optical $V$, $R$ and $I$ bands from 2005 to 2016. Intraday variability (IDV) is detected on seven nights. The variability amplitudes for most of nights are less than 10\% and four nights more than 20\%. When considering the nights with time spans $>4$ hours, the value of duty cycle (DC) is 14.17 per cent. Over the twelve years, the overall magnitude and color index variabilities are $\bigtriangleup I=0^{\rm m}.67$, $\bigtriangleup R=0^{\rm m}.72$, $\bigtriangleup V=0^{\rm m}.68$, and $\bigtriangleup (V-R)=0^{\rm m}.25$ respectively. The largest clear IDV has an amplitude of 42\% over just 5.8 minutes and the weakest detected IDV is 5.4\% over 175 minutes. The BWB (bluer when brighter) chromatic trend is dominant for 3C 273 and appears at different flux levels on intraday timescales. The BWB trend exists for short-term timescales and intermediate-term timescales but different timescales have different correlations. There is no BWB trend for our whole time-series data sets. A significant anti-correlation between BWB trend and length of timescales is found. Combining with $V$-band data from previous works, we find a possible quasi-periodicity of $P=3918\pm1112$ days. The possible explanations for the observed variability, BWB chromatic trend and periodicity are discussed.

\end{abstract}


\section{INTRODUCTION}

Active galactic nuclei (AGNs) are very energetic extragalactic sources powered by accretion on supermassive black hole (Esposito et al. 2015). The blazar subclasses' of radio-loud AGNs have jets pointing in the direction of the observer and are characterized by large amplitude and rapid variability at all wavelengths, high and variable polarization, superluminal jet speeds and compact radio emission (Angel \& Stockman 1980; Urry \& Padovani 1995). Blazars are often subclassified into BL Lacertae (BL Lac) objects and flat spectrum radio quasars (FSRQs). FSRQs have strong emission lines, while BL Lac objects have only very weak or non-existent emission lines. The emission of blazars is dominated by a relativistic jet which is boosted by beaming effect (Sandrinelli et al. 2016). The broadband spectral energy distributions (SEDs) of blazars have a double peaked structure. The low energy peak at the IR-optical-UV band is explained with the synchrotron emission of relativistic electrons, and the high energy peak at the GeV-TeV gamma-ray band is due to the inverse Compton (IC) scattering (e.g. Dermer et al. 1995; Dermer et al. 2002; Bottcher 2007). The hadronic model is an alternative explanation for the high energy emissions from blazars (e.g. Dermer et al. 2012).

Blazars show variability on different timescales from years down to
minutes (e.g. Poon et al. 2009; Zhang et al. 2008; Fan et al. 2005, 2009,
2014). Blazar variabilities can be broadly divided into intraday
variability (IDV) or micro-variability, short-term variability (STV)
and long-term variability (LTV). Variations in flux of a few tenths or hundredths
of a magnitude in a time scale of tens of minutes to a few hours are
often known as IDV (Wagner \& Witzel 1995). STV has time scales of
days to weeks, even months and LTV ranges from months to years
(Gupta et al. 2008; Dai et al. 2015). The IDV of blazars is the
least well understood type of variations but it can shed light on
the location, size, structure and dynamics of the emitting regions
and radiation mechanism (e.g. Wagner \& Witzel 1995; Ciprini et al.
2003, 2007; Kalita et al. 2015; Dai et al. 2015). The discovery of a periodicity in the
variability could have profound consequences in the global
understanding of the sources, constituting a basic block for
models (Sandrinelli et al. 2016). In order to search for periodic variations on many timescales, long-term observations are needed.

The quasar 3C 273 (R.A.=$12^{\rm h}29^{\rm m}06^{\rm s}.7$, decl.=$02^{\rm \circ}03^{\rm '}09^{\rm ''}$, J2000, redshift $z=0.158$) is the first quasar discovered by Smith \& Hoffleit (1963). It is classified as a bright gamma-loud FSRQ (Abdo et al. 2010a), and has been observed for more than 50 years in the radio band and more than 100 years in the optical band (Vol'vach et al 2013). The quasar has blazar-like (superluminal jet and high variability) and Seyfert-like (the strong blue bump, the soft X-ray excess, a weak and neutral iron line and variable emission lines) characteristics (Esposito et al. 2015; Chidiac et al. 2016). The flux and spectral variations of 3C 273 have been extensively studied over the entire electromagnetic spectrum (e.g. Xie et al. 1999, 2004a;
Ghosh et al. 2000; Mantovani et al. 2000; Collmar et al.
2000; Sambruna et al. 2001; Greve et al. 2002; Kataoka et
al. 2002; Courvoisier et al. 2003; Jester et al. 2005; Attridge
et al. 2005; Savolainen et al. 2006; McHardy et al. 2007;
Soldi et al. 2008; Espaillat et al. 2008; Pacciani et al. 2009; Dai et al. 2009; Fan
et al. 2009, 2014; Abdo et al. 2010; Ikejiri et al. 2011; Rani et al. 2011; Vol'vach et al. 2013; Beaklini \& Abraham
2014; Kalita et al. 2015; Esposito et al. 2015; Madsen et al. 2015; Chidiac et al. 2016). Soldi et al. (2008) studied the multiwavelength variability of 3C 273. Their results showed variability at all frequencies, with amplitudes and time scales strongly depending on the energy, and implied that either two separate components (possibly a Seyfert-like and a blazar-like) or at least two parameters with distinct timing properties account for the X-ray emission below and above $\sim$20 keV. Soldi et al. (2008) also presented that the optical component is consistent with optically thin synchrotron radiation from
the base of the jet and the hard X-rays would be produced through inverse Compton processes (SSC and/or EC) by the same electron population. Abdo et al. (2010) presented the $\gamma$-ray outburts light curves and spectral data from 3C 273 in September 2009. During these flares, 3C 273 reached very high brightness levels for quite short time intervals of $1-10$ days, and showed that the rise and the decay are asymmetric on timescales of 6 hr, and that the spectral index is significantly harder during the flares. Esposito et al. (2015) studied the high energy spectrum of 3C 273 and suggested a two-component model to explain the complete high energy spectrum. Chidiac et al. (2016) found that the variations at higher frequencies are leading the lower frequencies, which could be expected in the standard shock-in-jet model. Kalita et al. (2015) found high flux variability over long time scales in optical, UV and X-ray bands, and that on intraday time scales 3C 273 shows small amplitude variability in X-ray bands. Madsen et al. (2015) investigated the spectral variability in the NuSTAR band and found an inverse correlation between flux and photon index. In the optical band, unlike other blazars which are highly polarised, 3C 273 is a low-polarization quasar (Valtaoja et al. 1991; Xie et al. 1999; Chidiac et al. 2016). Generally, the properties
of 3C 273 are very different from that of BL Lac objects and highly polarized
quasars in the optical bands. It showed moderate variation in
amplitude. Variations of $0.2-0.3$ mag can be seen on a timescale of a
few days, but $\triangle V\geq1$ mag was not seen even in a
timescale of many years (Smith et al. 1987; Xie et al. 1999). Small amplitude changes
over a short time were reported (e.g. Moles et al. 1986; Rani et al. 2011). However, for the quasar, Fan et al. (2009) also presented intra-day variations of 0.55 mag in 13 minutes and 0.81 mag in 116 minutes. Paltani et al. (1998) presented two distinct variable components in the optical-ultraviolet bands. One rapidly variable component could be
from the accretion disk (Shields 1978; Soldi et al. 2008), or illumination
by an X-ray source (Ross \& Fabian 1993) or a hot
corona (Haardt et al. 1994). The second component seems to be related to synchrotron emission from the jet (Paltani et al. 1998). Optical variability is often associated with
color/spectral behavior in blazars, which can be used to
explore the emission mechanism (e.g. Gu \& Ai 2011; Zheng et al. 2008; Xiong et al. 2016). The results from Dai et al. (2009) implied that the spectrum becomes bluer (flatter) when the source becomes
brighter both on intraday variability and long-term variability for 3C 273. Ikejiri et al. (2011) found that 3C 273 exhibits ``bluer-when-brighter'' trends in their whole time-series data sets. Fan et al. (2014) found that the spectrum becomes flat when the source becomes bright, and that the emission in the optical bands consists of two components. The complexity of 3C 273 emission is due to the presence of different emission components. Therefore, long-term monitoring of 3C 273 is very important to better understand the physics of the quasar.

In view of these facts, we monitored the source in optical band from 2005 to 2016. We analyze the variability and spectral properties of 3C 273. Combining with $V$ band data from Soldi et al. (2008) and Fan et al. (2014), we search the periodic signals of 3C 273 in 48 years time spans. This paper is organized as follows. The observations and data analysis are described in Section 2. Section 3 presents the results. Discussion and conclusions are reported in Section 4. A summary is given in Section 5.

\section{OBSERVATIONS AND DATA ANALYSIS}

Our optical monitoring program of 3C 273 was carried out using the
2.4 m and 1.02 m optical telescopes located at the Yunnan Astronomical
Observatories (YAO) of China. The 2.4 m optical telescope is equipped
with two photometric detectors: PI VersArry 1300B CCD camera (PICCD) and Yunnan Faint Object Spectrograph and Camera (YFOSC)\footnote{http://www.gmg.org.cn}. The 2.4 m optical telescope used PICCD from 2008 to 2012 but YFOSC after 2012. The PICCD with 1340 $\times$ 1300 pixels covers a field of view (FOV) 4.48 $\times$ 4.40 arcmin$^2$ at the Cassegrain focus. The readout noise and gain are 6.05 electrons and 1.1 electrons/ADU, respectively. YFOSC has a FOV of about 10 $\times$ 10 arcmin$^2$ and 2000 $\times$ 2000 pixels for photometric observation. Each pixel corresponds to 0.283 arcsec/pixel. The readout noise and gain of the YFOSC CCD are 7.5 electrons and 0.33 electrons/ADU, respectively (Liao et al. 2014). For 1.02 m optical telescope\footnote{http://1m.ynao.ac.cn}, prior to 2006, the entire CCD chip (1024 $\times$ 1024 pixels) covered ~6.5 $\times$ 6.5 arcmin$^2$. The readout noise and gain of the CCD are 3.9 electrons and 4.0 electrons/ADU, respectively. After 2009, the telescope was equipped with an Andor DW436 CCD (2048 $\times$ 2048 pixels) camera at the Cassergrain focus ($f=13.3$~m), with pixel size 13.5 $\times$ 13.5 $\mu m^2$. The readout noise and gain are 6.33 electrons and 2.0 electrons/ADU, respectively, with 2 $\mu$s (readout time per pixel) or 2.29 electrons and 1.4 electrons/ADU with 16 $\mu$s. The FOV of the CCD image is 7.3 $\times$ 7.3 arcmin$^2$ and its pixel scale is 0.21 arcsec/pixel (Dai et al. 2015; Xiong et al. 2016). The standard Johnson broadband filters were used for the two telescopes.

Our photometry observations were performed in $V$, $R$ and $I$ bands through two modes. The first
mode was that all of the observations were completed for same optical band and then moved to the next band. The other was a cyclic mode among $V$, $R$ and $I$ bands. The optical observations in the $V$, $R$ and $I$ bands were in a corresponding cyclic mode for half of the total data. The exposure times from 0.17 to 8.3 minutes were chosen according to different seeing and brightness of the source. Moreover, in order to search for very fast variability ($<10$ minutes) and obtain more data, the short exposure time in the first mode was adopted. For same optical band in our observatories, the time resolutions (minimum times between two adjacent data points for same optical band) were from 0.2 $-$ 20.5 minutes and most of time resolutions $<$ 15 minutes.  Therefore, we can consider these data in a cyclic mode as quasi-simultaneous measurements ($<$ 20 minutes). The rest of data were not explored for analyzing inter-band time lag and color index because their $V$, $R$ and $I$ bands were not observed in a corresponding cyclic mode. The observation log is given in Table 1 where we have listed observation date, time spans, time resolutions and number of data points for each date in different bands. From Table 1, there are many nights with time spans $<3$ hours due to the nights with bad weather and those devoted to other targets.

After correcting flat-field, bias, aperture photometry was performed using the APPHOT task of
IRAF\footnote{\scriptsize{IRAF is distributed by the National Optical
Astronomy Observatories, which are operated by the Association of
Universities for Research in Astronomy, Inc., under cooperative
agreement with the National Science Foundation.}}. The aperture
radius of 2$\times$FWHM was selected, considering the best S/N ratio. The finding chart of 3C 273 was obtained from the webpage\footnote{https://www.lsw.uni-heidelberg.de/projects/extragalactic/charts/1226+023.html}. Besides instrumental magnitude of 3C 273, we measured instrumental magnitudes of 4 comparison stars (C, D, E and G in the finding chart) on the same field. We chose C and D as comparison stars because C was the brightest comparison star and the differential magnitude between C and D was the smallest variations among the 4 comparison stars. Following Zhang et al. (2004, 2008), Fan et al. (2014), Bai et al. (1998), the source magnitude was given as the average of the values derived with respect to the two comparison stars ($\frac{m_C+m_D}{2}$, $m_C$ is the blazar magnitude obtained from standard star C and $m_D$ from standard star D). The magnitudes of comparison stars in the field of
3C 273 were taken from Smith et al. (1985). The rms errors
of the photometry on a certain night are calculated from the
two comparison stars, star C and star D, in the usual way:
\begin{equation}
\sigma=\sqrt{\sum \frac{(m_i-\overline{m})^2}{N-1}},~~~~i=1,2,3,...,N,
\end{equation}
where $m_i = (m_{\rm C}-m_{\rm D})_{i}$ is the differential
magnitude of stars $C$ and $D$, while
$\overline{m}=\overline{m_{C}-m_{D}}$ is the averaged
differential magnitude over the entire data set, and $N$ is the
number of the observations on a given night. In order to further
quantify the reliability of variability, the value of $S_x$ can be calculated as (e.g. Hu et al. 2014)
\begin{equation}
S_x=m_i-\overline{m},~~~x=V,R,I
\end{equation}
where $m_i$ and $\overline{m}$ are same with equation (1).

The variability amplitude (Amp) can be calculated by (Heidt \&
Wagner 1996)
\begin{equation}
{\rm Amp}=100\times \sqrt{(A_{max}-A_{min})^2-2\sigma^2}~{\rm per
cent},
\end{equation}
where $A_{max}$ and $A_{min}$ are the maximum and minimum magnitude,
respectively, of the light curve for the night being considered,
and $\sigma$ is rms errors.

The duty cycle (DC) is calculated as (Romero et al. 1999; Stalin et
al. 2009; Hu et ai. 2014)
\begin{equation}
{\rm DC}=100\frac{\sum^n_{i=1}N_i(1/\triangle
T_i)}{\sum^n_{i=1}(1/\triangle T_i)}{\rm per~cent},
\end{equation}
where $\triangle T_i=\triangle T_{i,obs}(1+z)^{-1}$, $z$ is the
redshift of the object and $\triangle T_{i,obs}$ is the duration of
the monitoring session of the \emph{ith} night. Note that since for
a given source the monitoring durations on different nights were not
always equal, the computation of DC has been weighted by the actual
monitoring duration $\triangle T_i$ on the \emph{ith} night. $N_i$
will be set to 1 if intraday variability is detected, otherwise
$N_i=0$ (Goyal et al. 2013).

The actual number of observations for 3C 273 is 105 nights obtaining 1901 $I$-band, 1707 $R$-band and 1544 $V$-band data points. The results of observations are given in Table $2-4$ for filters $I$, $R$ and $V$.

\section{RESULTS} \label{bozomath}
\subsection{Variability}

A number of statical tests have been proposed to prove the validity of IDV
reports. Normally, the $C$-test was often used to analyze the IDV.
However, de Diego (2010) has found that the $C$-test is not a
reliable methodology because it does not have a Gaussian
distribution and its criterion is too conservative. At present, the
optical IDV is often analyzed by the $F$-test, $\chi^2$-test, modified
$C$-test and one-way analysis of variance (de Diego 2010; Joshi et
al. 2011; Goyal et al. 2012, 2013; Hu et al. 2014; Dai et al. 2015;
Agarwal \& Gupta 2015). Romero et al. (1999) introduced the
variability parameter, $C$, as the average value between $C_1$ and
$C_2$:
\begin{equation}
C_1=\frac{\sigma(BL-StarA)}{\sigma(StarA-StarB)},
C_2=\frac{\sigma(BL-StarB)}{\sigma(StarA-StarB)},
\end{equation}
where (BL-StarA), (BL-StarB) and (StarA-StarB) are the differential
instrumental magnitudes of the blazar and comparison star A, the
blazar and comparison star B, and comparison star A and B. $\sigma$
is the standard deviation of the differential instrumental
magnitudes. The adopted variability criterion requires $C\geq2.576$,
which corresponds to a 99 per cent confidence level. Despite the
very common use of the $C$-statistics, de Diego (2010) has pointed
out that it has severe problems.

The $F$ test is thought to be a proper statistics to
quantify variability (e.g. de Diego 2010; Joshi et al. 2011; Goyal
et al. 2012; Hu et al. 2014; Agarwal \& Gupta 2015; Xiong et al. 2016). $F$ value is
calculated as
\begin{equation}
F_1=\frac{Var(BL-StarA)}{Var(StarA-StarB)},
F_2=\frac{Var(BL-StarB)}{Var(StarA-StarB)},
\end{equation}
where Var(BL-StarA), Var(BL-StarB) and Var(StarA-StarB) are the
variances of differential instrumental magnitudes. The $F$ value
from the average of $F_1$ and $F_2$ is compared with the critical
$F$-value, $F^\alpha_{\nu_{bl},\nu_\ast}$, where $\nu_{bl}$ and
$\nu_{\ast}$ are the number of degrees of freedom for the blazar and
comparison star respectively ($\nu=N-1$), and $\alpha$ is the
significance level set as 0.01 (2.6$\sigma$). If the average $F$
value is larger than the critical value, the blazar is variable at a
confidence level of 99 per cent. Other alternatives to the standard
$F$-test are the use of one-way analysis of variance (ANOVA) or $\chi^2$ test (e.g. de Diego 2010). de Diego (2010) presented that the
ANOVA is powerful and robust estimator for microvariations. We use the ANOVA in our analysis because
it does not rely on error measurement but derives the expected variance from
subsamples of the data. Considering the time of exposure, we bin the data in a group of three or five observations (see Xiong et al. 2016 and de Diego 2010 for detail). This method is used only for light curves with more than 20
observations on a given night, but if the measurements in the last group are less than 3 or 5, then it is merged with the previous group. The critical value of ANOVA can be obtained by $F^\alpha_{\nu_1,\nu_2}$ in the $F$-statistics, where $\nu_1=k-1$ ($k$ is the number of groups), $\nu_2=N-k$ ($N$ is the number of measurements) and $\alpha$ is the significance level (Hu et al. 2014). When analyzing, we remove the outliers that $S_x$ is more than 3$\sigma$.

The blazar is considered as variability (V) if the light curve satisfies the two criteria of $F$-test and ANOVA. The blazar is considered as probably variable (PV) if only one of the above two criteria is satisfied. The blazar is considered as non-variable (N) if none of the criteria are met. We only analyze the data with exposure times of more than 1 minute because much shorter exposure times produce lower quality data and their use makes an analysis of detection of IDV much more difficult. Our results of analysis on intraday variability are shown in Table 5. From Table 5, it can be seen that there is IDV found on five nights ($I$ bands on 2013 April 04, 2015 April 15, 2010 March 19, 2009 January 17 and $V$ band on 2016 April 25; Fig. 1). As an illustration, the specific variability on two nights are presented. On 2013 April 04, the largest magnitude change of $I$ band is $\bigtriangleup I=0^{\rm m}.055$ in 175 minutes from MJD=56386.621 to MJD=56386.742 corresponding to the variability amplitude Amp=5.39\%. On 2015 April 15,  3C 273 brightens by $\bigtriangleup I=0^{\rm m}.038$ in 52 minutes from the beginning of MJD=57127.656 to MJD=57127.692, and then fades by $\bigtriangleup I=0^{\rm m}.029$ in 41 minutes from MJD=57127.692 to MJD=57127.721. From Table 5 and Fig. 1, we can see that the variability amplitudes on these nights are close to $5\%$. For $R$ band from 2014 April 25 and $I$ band from 2015 May 17, the light curves meet the criterion of $F$-test, but can not reach the critical value of ANOVA. Though we use different bin sizes ($N=2-6$), the two nights still can not reach the critical value of ANOVA. It is can be seen that the light curves of the two nights have large variation with Amp$>$30\% (Table 5 and Fig. 1). ANOVA test is used to compare the means of a number of samples. For light curves of the two nights, when 3C 273 occurs flare or darkens, there are only a few data points in the corresponding time interval which reduces the variations between groups, and then causes that the ANOVA test can not detect the IDV (see de Diego 2010 for detail). Therefore, according to results from $F$-test, the two nights are detected as IDV with large variability amplitude. On 2014 April 25, 3C 273 brightens by $\bigtriangleup R=0^{\rm m}.27$ in 51 minutes from the beginning of MJD=56772.668 to MJD=56772.703, and then quickly fades by $\bigtriangleup R=0^{\rm m}.27$ in 26 minutes from MJD=56772.703 to MJD=56772.721. On 2015 May 17, 3C 273 brightens by $\bigtriangleup I=0^{\rm m}.27$ in 18 minutes from the beginning of MJD=57159.651 to MJD=57159.663, and then quickly fades by $\bigtriangleup I=0^{\rm m}.42$ in 5.8 minutes from MJD=57159.663 to MJD=57159.667 and again brightens by $\bigtriangleup I=0^{\rm m}.22$ in 5.9 minutes from the beginning of MJD=57159.667 to MJD=57159.671. The outliers ($S_x>3\sigma$) only appear in two nights (2015 April 15 and 2016 April 25) and never exceed two data points for per night.

To sum up, IDV is detected on seven nights. For the seven nights, we also check the color variations on intraday timescales. However, the results from two statistical tests do not show the corresponding color variations on intraday timescales. There are fourteen nights detected as PV excluding the seven nights. When estimating the variability amplitude (Amp), we only consider the night detected as variability and possible variability. The distributions of variability amplitude from different bands are given in Fig. 2 which shows that the variability amplitudes for most of nights are less than $10\%$ and four nights more than $20\%$ (also see Table 5). The correlations between variability amplitudes and the source average brightness are shown in Fig. 3. The results from the analysis of Spearman rank indicate that there are not significant correlations between variability amplitude and brightness (significance level $P>0.05$, $N=$11, 7, 9 for $I$, $R$ and $V$ bands). However, there are not enough data to support the results. Making use of Equation (4), we calculate DC of intraday variability. The value of DC is 10.84 per cent for V case and 53.65 per cent for PV+V cases. When considering the nights with time spans $>4$ hours, the value of DC is 14.17 per cent for V case and 55.24 per cent for PV+V cases.

Long-term light curves and color index variations are given in Fig. 4. From Fig. 4, we can see that on the whole the quasar becomes dark from 2005 to 2015 but also bright between different years; in 2015, the quasar reaches a low flux state; compared with 2015, the quasar in 2016 has a tendency to brighten; there is a color index variability for long timescales. Over the twelve years, the overall magnitude and color index variabilities are $\bigtriangleup I=0^{\rm m}.67$, $\bigtriangleup
R=0^{\rm m}.72$, $\bigtriangleup V=0^{\rm m}.68$, $\bigtriangleup
(V-R)=0^{\rm m}.25$ respectively. The magnitude distributions in the $V$, $R$ and $I$ bands are $13^{\rm m}.10<V<12^{\rm m}.42$, $13^{\rm m}.09<R<12^{\rm m}.37$ and $12^{\rm m}.59<I<11^{\rm m}.93$ respectively. The average values of magnitude and color index are $<I>=12^{\rm m}.178\pm0^{\rm m}.118$, $<R>=12^{\rm m}.645\pm0^{\rm m}.107$, $<V>=12^{\rm m}.791\pm0^{\rm m}.106$ and $<V-R>=0^{\rm m}.126\pm0^{\rm m}.023$ respectively.

\subsection{Cross-correlation analysis and variability timescales}

Using the $z$-transformed discrete correlation function (ZDCF; Alexander 1997), we perform the inter-band correlation analysis and search for the possible inter-band time delay. The ZDCF estimates the cross-correlation function (CCF) in the case of non-uniformly sampled light curves, and deals with sparsely and unequally sampled light curves better than both the interpolated cross-correlation function (ICCF) and the discrete correlation function (DCF; Alexander 1997; Edelson et al. 1996; Giveon et al. 1999; Roy et al. 2000). The ZDCF attempts to correct the biases that affect the original DCF by using equal-population binning. The ZDCF involves three steps (Alexander 1997; Oscoz et al. 2001). (i) All possible pairs of observations are sorted according to their time-lag, and binned into equal population bins of at least 11 pairs. Multiple occurrences of the same point in a bin are discarded so that each point appears only once per bin. (ii) Each bin is assigned its mean time-lag and the intervals above and below the mean that
contain $1\sigma$ of the points each. (iii) The correlation coefficients of the bins are calculated and $z$-transformed. The error is calculated in $z$-space and transformed back to $r$-space. The time-lag corresponding to the maximum value of the ZDCF is assumed as the time delay between both components. The ZDCF code of Alexander et al. (1997) can automatically set how many bins are given and used to calculate the inter-band correlation and the ACF. For each correlation, a Gaussian fitting is made to find the central ZDCF points. The time where the Gaussian profile peaks denotes the lag between the correlated light curves (Wu et al. 2012). The results that can determine the time delay are displayed in Fig. 5. From Fig. 5 and results of Gaussian fitting, the corresponding time lags are $-0.002\pm0.001,
0.002\pm0.002, -0.007\pm0.003, 0.003\pm0.002, 0.011\pm0.004, 0.014\pm0.003,
0.026\pm0.006$ and $-0.005\pm0.004$ respectively. Considering time resolutions, we can not find significant time lags between the $V$ band magnitude and $I$ band magnitude. For the rest of nights, there are not enough data points or good Gauss profile to determine time lags.

The autocorrelation function (ACF) is defined by
\begin{equation}
ACF(\tau)\equiv\langle(m(t)-\langle m\rangle)\cdot(m(t+\tau)-\langle m\rangle)\rangle
\end{equation}
where brackets denote a time average. The ACF measures the correlation of the lightcurve
with itself, shifted in time, as a function of the time lag $\tau$ (Giveon et al. 1999). The zero-crossing time is the shortest time it takes the ACF to fall to zero (Alexander 1997). If there is an underlying signal in the
light curve, with a typical timescale, then the width of the ACF
peak near zero time lag will be proportional to this timescale
(Giveon et al. 1999; Liu et al. 2008). The zero-crossing time is a well-defined quantity and
used as a characteristic variability timescale (e.g. Alexander 1997; Giveon et al. 1999; Netzer et al. 1996; Liu et al. 2008; Liao et al. 2014). The width of the ACF may be related to a
characteristic size scale of the corresponding emission region (Abdo et al. 2010b; Chatterjee et al. 2012). Another function used in variability studies to
estimate the variability timescales is the first-order structure function (SF; e.g. Trevese et al. 1994) defined by
\begin{equation}
SF(\tau)\equiv\sqrt{\langle(m(t+\tau)-m(t))^2\rangle}.
\end{equation}
There is a simple relation between the ACF and the SF,
\begin{equation}
SF^2(\tau)=2(V-ACF(\tau)),
\end{equation}
where $V$ is the variance of the lightcurve (Giveon et al. 1999). We therefore perform only an ACF analysis on
our lightcurves. The ACF was estimated by ZDCF. We only analyze the nights detected as intraday variability. The results used to estimate characteristic
variability timescales are given in Fig. 6. Following Giveon et al. (1999), Liu et al. (2008) and Liao et al. (2014), we use a least-squares procedure to fit
a fifth-order polynomial to the ACF, with the constraint that ACF($\tau=0$)=1. The fitting results show that the detected variability timescales are 0.107, 0.023, 0.005 and 0.028 days for $I$ band on 2013 April 04, $R$ band on 2014 April 25, $I$ bands on 2015 May 17 and 2015 April 15. In Section 3.1, we have given the timescales corresponding to the change of brightness on these nights (also see Fig. 1). By comparison, we can obtain that the variability timescales from ACF analysis are consistent with the timescales corresponding to the change of brightness.

\subsection{Correlation between Magnitude and Color Index}

In this Section, we analyze the relationships between magnitude and color index for intraday timescales,
short-term timescales, and whole time-series data sets. The whole time-series data sets include data from 2005 to 2016, but exclude the data from 2006 to 2008 due to non quasi-simultaneous measurements. For the color index, we correct the Galactic extinction. The correction factors of Galactic extinction are from the Schlafly \& Finkbeiner (2011). We concentrate on $V-R$ index and $V$ magnitude because $V-R$ index versus $V$ magnitude is frequently studied. When exploring the relationships, we only analyze the data with more than 9 color indices obtained on intraday timescales and quasi-simultaneous measurements. The results of correlations between $V-R$ index and $V$ magnitude are given in Table 6. As an example, Fig. 7 shows the correlations between $V-R$ index and $V$ magnitude on intraday timescales. The results from Table 6 show that most of nights have strong correlations between $V-R$ index and $V$ magnitude on intraday timescales, and the rest of nights moderate or weak correlations. Therefore a bluer-when-brighter (BWB) chromatic trend is dominant for 3C 273 on intraday timescales. The BWB trend exists for short-term timescales and intermediate-term timescales but different timescales have different correlations between $V-R$ index and $V$ magnitude (Table 6). The correlation between $V-R$ index and $V$ magnitude in whole time-series data sets is shown in Fig. 8. The result of correlation analysis from Table 6 shows that there is no BWB trend in whole time-series data sets. We have the opportunity to explore the relationship between BWB trend and length of timescales in $V$-band because of long-term optical monitoring. We use the coefficient of correlation from errors weighted linear regression analysis to indicate the intensity of BWB trend, and the sum of monitoring time spans on
intraday timescales to indicate length of timescales. The analysis of Spearman rank shows that there is significant anti-correlation between BWB trend and length of timescales ($r=-0.491, P=0.002$; see Fig. 9).

\subsection{Periodicity Analysis}

In order to search for periodicity, well-covered long-term light curves are required. Soldi1 et al. (2008) presented an update of the 3C 273's database hosted by the ISDC\footnote{http://isdc.unige.ch/3c273/}. The 3C 273's database which was first published by Turler et al. (1999) is one of the most complete multiwavelength databases currently available for the quasar. The time spans of the optical $V$ band database are from 1968 to 2005. However, there are no the $V$ band data from 1998 to 1999. We also compiled $V$ band data in $1998-1999$ from Fan et al. (2014). Our data from 2005 to 2016 are update and supplementary for above data. For the data from Fan et al. (2014) and the present work, we convert magnitudes to fluxes in the Geneva photometric system as described in Turler et al. (1999), without any additional correction (Vcorr = 0) because the data from Fan et al. (2014) and the present work used standard Johnson photometric system while Soldi et al. (2008) and Turler et al. (1999) used Geneva photometric system. For data from Soldi et al. (2008), we consider data of Flag=0 in order to use only the best quality data and exclude the data with the influence of strong synchrotron flares (Flag=1). Combining with all data, we search the periodic signals of 3C 273 in 48 years time spans. The long-term light curves (bin=1 day) are given in Fig. 10. When having more than one data on intraday timescales, we use standard deviation as error.

For periodicity analysis of AGNs, unevenly sampled data, frequency dependent red noise, flares of high activity and total monitoring time need to be considered (e.g. Sandrinelli et al. 2016; Fan et al. 2014; Vaughan 2005, 2010; Schulz \& Statteger 1997; Schulz \& Mudelsee 2002). For the data, the strong synchrotron flares are excluded. The time spans from the data are 48 years. Therefore, when analyzing periodicity, we mainly consider unevenly sampled data and red noise. In addition, in order to reduce false periodicity produced by short time scale variability, we pay more attention to periodicity in year timescales. In order to increase the reliability of periodicity analysis, we use three methods (REDFIT: Schulz \& Mudelsee 2002; Lomb-Scargle method: Lomb 1976 and Scargle 1982; Jurkevich method: Jurkevich 1971) to explore periodicity. The periodicity can be considered valid only if three methods have consistent results. Schulz \& Mudelsee (2002) presented a computer program (REDFIT3.8e\footnote{http://www.geo.uni-bremen.de/geomod/staff/mschulz/}) estimating red-noise spectra directly from unevenly spaced time series by fitting a first-order autoregressive (AR1) process. The program can be used to test if peaks in the spectrum of a time series are significant against the red-noise background from an AR1 process, and removes the bias of this Fourier transform for unevenly spaced data by correcting for the effect of correlation between Lomb-Scargle Fourier components. However, this program has two underlying assumptions: (i) the noise background recorded in a time series can indeed be approximated by an AR1 process; (ii) the distribution of data points along the time axis is not too clustered. For our data, the results of non-parametric runs test indicate that the spectrum is consistent with an AR1 model (rtest=139 falls inside 98\% acceptance region [129, 170]; see Schulz \& Mudel 2002 and usage38e for further detail). The Fig. 10 also shows that the distribution of data points along the time axis is not too clustered. So it is appropriate to use REDFIT3.8e for periodicity analysis. The Lomb-Scargle method is commonly used to detect periodicity in unevenly sampled time series. The periodogram is a function of circular frequency $\omega$, and is defined by the formula (Li et al. 2016)

\begin{equation}
 P_{X}(\omega)\equiv \frac{1}{2}\{\frac{[\Sigma _{i}X(t_{i}){\rm cos}\omega(t_{i}-\tau)]^{2}}{\Sigma_{i}{\rm cos}^{2}\omega (t_{i}-\tau)}+\frac{[\Sigma _{i}X(t_{i}){\rm sin}\omega(t_{i}-\tau)]^{2}}{\Sigma_{i}{\rm sin}^{2}\omega
 (t_{i}-\tau)}\},
\label{eq:LebsequeLS1}
\end{equation}
where $X(t_{i})$ ($i=0, 1...,N_{0}$) is a time series. $\tau$ is calculated by the equation
\begin{equation}
 \tau=\frac{1}{2\omega} {\rm tan}^{-1}[\frac{\Sigma _{i}{\rm sin}2\omega t_{i}}{\Sigma _{i}{\rm cos}2\omega
 t_{i}}],
\label{eq:LebsequeLS2}
\end{equation}
where $\omega=2\pi\nu$. Thus, the periodogram is a function of
frequency $\nu$.
For a true signal $X(t_{i})$, the power in $P_{X}(\omega)$ would present a peak, or the power of a purely noise signal would be a
exponential distribution. For a power level $z$, the False Alarm Probability (FAP) is calculated by (Scargle 1982; Press et al. 1994; Li et al. 2016)
\begin{equation}
 p(>z)\approx N\cdot {\rm exp}(-z),
\label{eq:LebsequeIp}
\end{equation}
where $N$ is the number
of data point. The Jurkevich method is an alternative and powerful method which is insensitive to the mean shape of periodicity, less sensitive than Fourier analysis to uneven sampling. The Jurkevich method involves testing a run of trial periods around which the data is folded. The data is binned in phase around each trial period and the total variance for all the points in the bin computed. A good period will give a much reduced variance ($V_\textrm{m}^\textrm{2}$). No firm rule exists for assessing the significance of a minimum in the $V_\textrm{m}^\textrm{2}$. But a good guide is the fractional reduction of the variance ($f=\frac{1-V_\textrm{m}^\textrm{2}}{V_\textrm{m}^\textrm{2}}$). Generally, a value $f\geq0.5$ ($V_\textrm{m}^\textrm{2}\leq0.667$) implies that there is a very strong periodicity in the data (Jurkevich 1971; Kidger et al. 1992). We use half of full width at half-maximun (FWHM) of peak as errors of periodicity. For REDFIT, when estimating errors of periodicity. we choose the minimum of the peak height above the average power level of red-noise in the vicinity of the investigated period. The results of periodicity analysis are given in Fig. 11. The results of REDIFIT show that there is a broad peak (the center frequency $V_{\rm peak}=2.55\times10^{-4}$ day$^{-1}$) with 90\% significance levels and above the red-noise power level. The corresponding periodicity is $3918\pm1112$ days in year timescales. The results of Lomb-Scargle method show that there are three significant peaks $P_1$ ($4961\pm1072$ days), $P_2$ ($3848\pm471$ days) and $P_3$ ($2793\pm325$ days) above 0.01 FAP levels. The results of Jurkevich method show that from 2749 ($\pm36$) days to 5594 ($\pm44$) days there are some strong periodicity signals. The strong periodicity signal of $3950\pm14$ days also appears in the results of Jurkevich method. We do not consider periods $>5840$ days because the periods (considered) are limited to three times periods less than total time spans (48 years). From above results, we should note that the results from REDFIT and Lomb-Scargle methods have large errors and that for Jurkevich method there are some possible periods with the range from 2749 days to 5594 days. Therefore, within the errors of periods, the periods estimated by the three methods have consistent results. A possible quasi-periodicity of $P=3918\pm1112$ days is found.

\section{DISCUSSION AND CONCLUSIONS} \label{bozomath}

\subsection{Variability}

Long-term multi-band observations of the quasar 3C 273 were performed from 2005 to 2016 in the $V$, $R$ and $I$ bands. The magnitude distributions in the $V$, $R$ and $I$ bands are $13^{\rm m}.10<V<12^{\rm m}.42$, $13^{\rm m}.09<R<12^{\rm m}.37$ and $12^{\rm m}.59<I<11^{\rm m}.93$ respectively. The average values of magnitude are $<I>=12^{\rm m}.178\pm0^{\rm m}.118$, $<R>=12^{\rm m}.645\pm0^{\rm m}.107$, $<V>=12^{\rm m}.791\pm0^{\rm m}.106$ respectively. Toone (2004) observed this
source during the period of $1980-2004$ and presented the yearly
averaged light curve. The magnitude distributions in $V$ band from
Toone (2004) are $13^{\rm m}.12-12^{\rm m}.5$ with $<V>=12^{\rm
m}.86\pm0^{\rm m}.13$. The magnitude distributions in $2000-2008$ from
Fan (2009) are $13^{\rm m}.567-12^{\rm m}.204$, $13^{\rm
m}.313-12^{\rm m}.014$ and $12^{\rm m}.669-11^{\rm m}.628$ for $V$,
$R$ and $I$ bands respectively. The magnitude distributions in
$1998-2008$ from Fan (2014) are $12^{\rm m}.9-12^{\rm m}.5$, $12^{\rm
m}.8-12^{\rm m}.3$ and $12^{\rm m}.2-11^{\rm m}.85$ for $V$, $R$ and
$I$ bands respectively. The magnitude distributions in $2003-2005$
from Dai (2009) are $12^{\rm m}.824-12^{\rm m}.663$, $12^{\rm
m}.714-12^{\rm m}.338$ and $12^{\rm m}.318-12^{\rm m}.055$
in $V$, $R$ and $I$ bands with the mean magnitudes $12^{\rm m}.698$,
$12^{\rm m}.441$ and $12^{\rm m}.139$ respectively. The magnitude
distributions in 1996 from Xie (1999) are $13^{\rm m}.9-12^{\rm
m}.86$, $12^{\rm m}.98-12^{\rm m}.6$ and $12^{\rm m}.47-12^{\rm m}.14$ in $V$, $R$ and $I$ bands respectively. Ikejiri et al. (2011) presented that the $V$-band magnitude distributions between 2008 and 2010 are $12^{\rm m}.75-12^{\rm m}.51$. From the above references, we obtain that the maximum and minimum magnitudes for this source data are $13^{\rm m}.9-12^{\rm m}.2$, $13^{\rm m}.31-12^{\rm m}.01$ and $12^{\rm m}.67-11^{\rm m}.63$ for the $V$, $R$ and $I$ bands respectively. Through comparisons, we can find that our distributive ranges of magnitude in the $V$, $R$ and $I$ bands are within the ranges of the maximum and minimum magnitudes from previous observations. From Fig. 10, we also get same conclusion. Moreover, our long-term multi-band observations are update and supplementary for the quasar data. During our observation, on the whole the quasar becomes dark from 2005 to 2015 but also bright between different years. In 2015, the quasar reaches a low flux state. Compared with 2015, the quasar in 2016 has a tendency to brighten. Over the twelve years, the overall magnitude variabilities are $\bigtriangleup I=0^{\rm m}.67$, $\bigtriangleup R=0^{\rm m}.72$, $\bigtriangleup V=0^{\rm m}.68$ respectively. These results indicate that unlike the other blazars, 3C 273 shows moderate magnitude variations in a timescale of many years.

In order to accurately determine the optical IDV, we use two statical tests to analyze the observation data. The blazar is considered as variability only if the light curve satisfies the two criteria of $F$-test and ANOVA. When analyzing, we remove the outliers that $S_x$ is more than $3\sigma$. Through these tests and processing, we attain a reliable result that there is IDV found on five nights. On 2013 April 04, the largest magnitude change of $I$ band is $\bigtriangleup I=0^{\rm m}.055$ in 175 minutes corresponding to the variability amplitude Amp=5.39\%. The variability amplitudes for the rest of the four nights are close to 5\%. Rani et al. (2011) presented two nights detected as IDV (Amp$\sim5\%$) by mean of $C$ and $F$ tests. For $R$ band from 2014 April 25 and $I$ bands from 2015 May 17, the light curves meet the criteria of $F$-test, but can not reach the critical value of ANOVA. We still consider the two nights detected as IDV because the light curves of the two nights have large variation with Amp$>30\%$ and ANOVA is easy to miss this kind of IDV that there are only a few data points in the time interval of flaring or darkening (see Section 3.1). On 2015 May 17, 3C 273 quickly fades by $\bigtriangleup I=0^{\rm m}.42$ in 5.8 minutes. Fan et al. (2009) presented that the quasar shows intraday variations of 0.55 mag in 13 minutes and 0.81 mag in 116 minutes. Our results also obtain that the variability amplitudes for most of nights are less than $10\%$ and four nights more than $20\%$. The value of DC is 10.84 per cent for V case and 53.65 per cent for PV+V cases. Extensive IDV studies of different subclasses of AGNs
revealed that the occurrence of IDV in a blazar observed on a timescale of $<6$ hours is $\sim60-65\%$ and if the blazar is observed more than 6 hours then the possibility of IDV detection is $80-85\%$ (Gupta \& Joshi 2005; Rani et al. 2011). Then it is quite likely that the value of DC is related with time spans of observation. When considering the nights with time spans $>4$ hours, the value of DC is 14.17 per cent for V case and 55.24 per cent for PV+V cases. Therefore, the quasar has low variability amplitudes and DC which is different from the other FSRQs. However, we still need more nights with longer time spans to confirm the results in the further. Although our results show that there are not significant correlations between variability amplitude and brightness, we do not consider this result as a reliable result because the result is not supported by enough data.

We notice that the variability amplitudes for the non-variable nights are non-zero and more than 5\% or even 10\% (Table 5). The variability amplitudes from three nights with exposure time $\sim$1 minute ($V$ bands on 2007 March 27, 2014 April 21 and 2016 April 24) are more than 5\%. If we choose the bin=3 minutes, the variability amplitudes decrease to 2.4\%, 6.17\% and 3.48\% respectively. Gopal-Krishna et al. (2003) and Goyal et al. (2013) have presented that IRAF and DAOPHOT produce nominal errors that are too small by factors of $\sim1.5$. It is possible that the non-zero variability amplitudes for the non-variable nights are due to too small values of $\sigma$ in Equation (3). Assuming variability amplitudes for the non-variable nights below 1\%, the ranges of underestimated factors are from 1 to 4.7 and average values $\sim2$. If the original variability amplitudes from Table 5 for the non-variable nights are below 1\%, the underestimated factors are set as 1, i.e. the value of $\sigma$ is not underestimated. For the nights detected as V case, using the underestimated factors $\sim2$, we re-calculate the variability amplitudes. The results show that compared with the original variability amplitudes from Table 5, the re-calculated variability amplitudes do not change significantly. Therefore, for some nights, the uncertainties in Table 2-4 are underestimated by factors of $\sim2$. On the other hand, the ANOVA (like $I$ band on 2015 May 17 or nights of low sampling rate) and standard $F$-test could fail to detect IDV (Joshi et al. 2011). This may also explain partly the problem.

From results of ACF, the detected characteristic variability timescales are 0.107, 0.023, 0.005 and 0.028 days for $I$ band on 2013 April 04, $R$ band on 2014 April 25, $I$ bands on 2015 May 17 and 2015 April 15. Given the minimum values of the lags that can be measured, we can not claim 0.023, 0.005 and 0.028 days as significance of timescales. Since the flares are essentially random and only 1 or 2 are seen on 2013 April 04, the variability timescale of 0.107 day (154 minutes) is considered as a possible variability timescale. We also use a least-squares procedure to fit fourth-order and sixth-order polynomials to the ACF on 2013 April 04. The fitting results of different order polynomials show that the differences of variability timescales are within 2 minutes.

For blazars, the intrinsic origin of IDV is due to relativistic jet activities or accretion disc instabilities (Marscher \& Gear 1985; Marscher et al. 2008; Chakrabarti \& Wiita 1993; Mangalam \& Wiita 1993). Extrinsic mechanisms include interstellar scintillation and Gravitational microlensing. Interstellar scintillation causes radio variations at low frequencies (Heeschen et al. 1987) and Gravitational microlensing is important on weeks to months time-scales (Agarwal \& Gupta 2015). Therefore, we do not consider extrinsic mechanisms to explain the origin of IDV. For 3C 273 in optical band, there is the blue-bump emission. Paltani et al. (1998) presented two possible variable components ($B$ and $R$) to explain the blue-bump emission. The $R$ component could be synchrotron emission not directly related to the radio-mm jet (Soldi et al. 2008). The $B$ component could be explained by reprocessing on an accretion disk (Paltani et al. 1998). The results from Paltani et al. (1998) obtained that the $B$ component has a maximum time scale of variability of about 2 years (a mean of 0.46 years) and the $R$ component has a longer time scale of variability. Then the two variable components can not explain the origin of IDV. In outburst state, the origin of IDV can be attributed to the shock-in-jet model: relativistic shocks propagating through a relativistic jet of plasma (Marscher \& Gear 1985; Marscher et al. 2008). The observed radiation is predominantly non-thermal Doppler boosted jet emission enhanced by that arising from shocks in the flows (Marscher \& Gear 1985). Models based on the instabilities on the accretion disk could yield IDV when the blazar is in the low state, because in the low state, any contribution from the jets is weak (Rani et al. 2011). For the quasar, the instabilities on the accretion disk could yield IDV in the modest state because it has obvious blue-bump emission. From Fig. 10, we get that on 2013 April 04 and 2014 April 25 the quasar has a mean flux level but on 2015 May 17 shows a low flux level. On 2015 May 17, the large variability amplitude (41.93\%) in 5.8 minutes is found. Although the quasar is in low flux level on this night, it is very impossible to use disc-based models to explain the type of variability because for luminous quasars with black hole masses in excess of $10^8 M_\odot$, even the fastest disc-based time-scale can be a few hours (Joshi et al. 2011). The type of variability is likely to have an association with jet activities. The shocks propagate down in relativistic jets, sweeping emitting regions. If the emitting regions have large intrinsic changes (magnetic field, particle velocity/distribution, a large number of new particle injected), and then we could see a large flare on a very short variability timescale. Another possible explanation is Doppler factor change of emitting region. The helical jet structure (Gopal-Krishna \& Wiita 1992) may cause the Doppler factor change in a very short variability timescale. On 2014 April 25, the large intrinsic changes in emitting regions are likely to explain the IDV. On 2013 April 04, the small variability amplitude ($\sim5\%$) in 154 minutes can be explained by turbulence behind an outgoing
shock along the jet (Agarwal \& Gupta 2015) or hotspots or disturbances in or above accretion discs (e.g. Chakrabarti \& Wiita 1993; Mangalam \& Wiita 1993; Gaur et al. 2012). The interpretations of IDV for the other nights are similar to that on 2013 April 04. If the observed 154 minutes indicate an innermost stable orbital
period from accretion disc, an upper limit can be obtained for the mass of the central
black hole (e.g. Xie et al. 2004b; Fan et al. 2014; Dai et al. 2015).
From Fan (2005) and Fan et al. (2014), the innermost stable orbit
depends on the black hole and the accretion disk, and $r\leq
c\triangle t_{\rm min}/(1+z)$, and then the upper limits of
black hole masses are (i) $M_8\leq1.2\times\frac{\triangle
t_{\rm min}(hr)}{1+z}$ for a thin accretion disk surrounding a
Schwarzschild black hole; (ii)
$M_8\leq1.8\times\frac{\triangle t_{\rm min}(hr)}{1+z}$ for a
thick accretion disk surrounding a Schwarzschild black hole; (iii)
$M_8\leq(\frac{7.3}{1+\sqrt{1-a^2}})(\frac{\triangle t_{\rm
min}(hr)}{1+z})$ for the case of a Kerr black hole, where $M_8$ is
the black hole mass in units of $10^8M_\odot$, and $a$ is an angular
momentum parameter. Then we obtain that the upper limits of black hole
mass are $M_8\leq2.66$ and $M_8\leq3.99$ for a
thin and thick accretion disks, and $M_8\leq16.18$ for the
extreme Kerr black hole case ($a=1$). The previous
reverberation mapping results obtained $M_8=2.3-5.5$ (Kaspi et al. 2000) and $M_8=65.9$ (Paltani \& Turler 2005). Therefore, based on this assumption that the observed 154 minutes indicate an innermost stable orbital period, our upper limits of black hole masses are consistent with reverberation mapping results.

Fan et al. (2009) found that the quasar shows the time delay between $V$ and $I$ bands. However, our results of ZDCF analysis indicate that there are not significant
time lags between the $V$ band magnitude and $I$ band magnitude. The optical inter-band time delay supports the shock-in-jet model. For our results, these factors (e.g. sampling rate, time length of observation, flat light curve) can cause that the significant time delay is not found.

\subsection{Relation of Color and Magnitude} \label{bozomath}

Over the twelve years, the overall color index variability is $\bigtriangleup
(V-R)=0^{\rm m}.25$. The average value of color index is $<V-R>=0^{\rm m}.126\pm0^{\rm m}.023$. From our results, the BWB chromatic trend is dominant for 3C 273 and appears at different flux levels for intraday timescales. The BWB trend exists for short-term timescales and intermediate-term timescales but different timescales have different correlations between $V-R$ index and $V$ magnitude. There is no BWB trend for our whole time-series data sets (12 years). The results from Dai et al. (2009) implied that the quasar has the BWB trend on both intraday
and long-term timescales (3 years). Ikejiri et al. (2011) found that 3C 273 exhibits the BWB trend in their whole time-series data sets (3 years). Fan et al. (2014) found that the spectrum becomes flat when the source becomes bright in an overall trend (about 5 years). However, there is a steeper with brighter trend in detail. When considering the $V$-band magnitude range from Dai et al. (2009), we still find the BWB chromatic trend in the long-term timescales for our data. When considering the $V$-band magnitude range from Ikejiri et al. (2011), we find a weak BWB chromatic trend for our data. The difference may be due to different samples and different color indices that Ikejiri et al. (2011) used $V-J$ while we used $V-R$. For Fig. 7 from Fan et al. (2014), there is a gap close to $V$-band flux $F_{\rm V}=28~ {\rm mJy}$. From our Fig. 10, we find that there are some data close to $V$-band flux $F_{\rm V}=28~ {\rm mJy}$. Then if the new data are included in Fig. 7 from Fan et al. (2014), the results of Fan et al. (2014) could change. In addition, we notice that compared with previous results, the time spans of our results are longer. The above discussions can explain differences between our results and previous results in the long-term timescales.

The BWB behaviour is most likely to support shock-in-jet model.  According to the shock-in-jet model, as the shock propagating down the jet strikes a region of high electron population, radiations at different visible colors are produced at different distances behind the shocks. High-energy photons from synchrotron mechanism typically emerge sooner and closer to the shock front than the lower frequency radiation thus causing color variations (Agarwal \& Gupta 2015). When we consider two distinct synchrotron components, the BWB trend could be explained if a flare component has a higher $V_{\rm peak}$ than an underlying component (Ikejiri et al. 2011). For 3C 273, there is a weak host galaxy contribution which is a non-variable redder component. Also the Doppler boosted jet emission almost invariably swamps the light from the host galaxy. Gravitational microlensing
is important on weeks to months timescales and achromatic. The optical emission may be contaminated by thermal emission from the accretion disk and the surrounding regions especially for quasars with the presence of blue bump. The quasar 3C 273 has a blue bump, which flattens
the spectral slope in the optical region. Since the thermal contribution is larger in the blue region, the composite spectrum would be flatter than the non-thermal component. Then, when the object is brightening, the non-thermal component has a more dominant contribution to the total flux, and the composite spectrum steepens (Gu et al. 2006). Consequently, in low flux level, the quasar could have redder with brighter (RWB) trend. For a very low flux state that the thermal radiation of disk dominates the total flux, accretion disc model can explain the BWB trend (Gu \& Li 2013; Li \& Cao 2008; Liu et al. 2016). So the BWB trend on intraday timescales is most likely explained by the shock-in-jet model, and also possible due to two distinct synchrotron components or accretion disc model. In addition, we find that there is a significant anti-correlation between BWB trend and length of timescales (Fig. 9). Sun et al. (2014) found that the color variability of quasars is
prominent on timescales as short as $\sim$10 days, but gradually reduces toward timescales up to years, i.e. the variable emission on shorter timescales is bluer than that on longer timescales. They proposed thermal accretion disk fluctuation model to explain the anti-correlation that fluctuations in the inner, hotter region of the disk are
responsible for short-term variations, while longer-term and stronger variations are expected from the larger and
cooler disk region. The BWB trend also can be interpreted in terms of two components: a `mild-chromatic' longer-term component and a `strongly-chromatic' shorter-term one, which can be likely due to Doppler factor variations on a ¡®convex¡¯ spectrum and intrinsic phenomena, respectively (Gu et al. 2006; Villata et al. 2002, 2004). The long-term achromatic trend could be due to superposition of different components (jet components, accretion disk components, Gravitational microlensing effect; Ikejiri et al. 2011; Bonning et al. 2012; Fan et al. 2008; Gaur et al. 2015; Agarwal \& Gupta 2015; Xiong et al. 2016).

\subsection{Periodicity} \label{bozomath}

In order to increase the reliability of periodicity analysis, we appropriately dealt with all of those factors: the best quality data excluding the influence of strong synchrotron flares, three methods for periodicity analysis, unevenly sampled data, red noise, periodicity in year timescales, three times periods less than total time spans. Converting magnitudes to fluxes in different photometric systems could bring into uncertainties which cause a small (false) variability in the light curve. In this case, if we search for periodicity on short or intraday timescales, the small variability could result in false periodicity. However, the small variability could not cause a significant impact on periodicity in year timescales because the periodicity mainly represents total change in year timescales. So analyzing periodicity in year timescales can reduce the effect from the uncertainties. By analyzing $V$-band data, a possible quasi-periodicity of $P=3918\pm1112$ days is found. However, it is very possible that a non-periodic model could provide a better fit. Fan et al. (2014) used $B$-band more than $100$ years data (from 1882 to 2012) to analyze the periods of the quasar. Their results obtained possible periods of $P = 21.10\pm0.14,
10.90\pm0.14, 13.20\pm0.09, 7.30\pm0.10, 2.10\pm0.06$ and
$0.68\pm0.05$ years for 3C 273. Smith \& Hoffleit (1963) suggested a period of $P = 12.7-15.2$ years for 3C 273. Babadzhanyants \& Belokon (1993) analyzed the $B$-band data of $1887-1991$ and found a period of $P = 13.4$ years. Vol'vach et al. (2013) obtained periods of $11.7\pm2.5,
7.2\pm0.8, 4.9\pm0.3$, and $2.8\pm0.3$ years based on the optical $B$-band light curve of $1963-2011$. Therefore, within the errors the quasi-period of $P=3918\pm1112$ days is consistent with the previous work. For the long-term period of 3C 273, some possible interpretations are as follows: the orbit of a perturbing object, precession of jet, a rotating helical jet structure and binary black hole model (Sandrinelli et al. 2016; Fan et al. 2014; Rani et al. 2009; Lehto \& Valtonen 1996; Marscher 2014; Villata \& Raiteri 1999; Vol'vach et al. 2013).

\section{SUMMARY}

We have monitored the quasar 3C 273 in the optical $V$, $R$ and $I$ bands from 2005 to 2016. In total, the actual number of observations for the quasar is 105 nights obtaining 1901 $I$-band, 1707 $R$-band and 1544 $V$-band data points, with rms error less than 0.06 mag. Our main results are the following:

(i) Intraday variability (IDV) is detected on seven nights. The variability amplitudes for most of nights are less than 10\% and four nights more than 20\%. When considering the nights with time spans $>4$ hours, the value of duty cycle (DC) is 14.17 per cent. Over the twelve years, on the whole the quasar becomes dark from 2005 to 2015 but also bright between different years. In 2015, the quasar reaches a low flux state. Compared with 2015, the quasar in 2016 has a tendency to brighten. The overall magnitude and color index variabilities are $\bigtriangleup I=0^{\rm m}.67$, $\bigtriangleup R=0^{\rm m}.72$, $\bigtriangleup V=0^{\rm m}.68$, $\bigtriangleup (V-R)=0^{\rm m}.25$ respectively.

(ii) The results of fitting ACF show a possible variability timescale of 0.107 day (154 minutes). The large
variability amplitude (41.93\%) in 5.8 minutes is found. The type of variability is likely to have an association with jet activities. The light curve of our observation shows the small variability amplitude ($\sim5\%$) in 154 minutes which can be explained by turbulence behind an outgoing shock along the jet or hotspots/disturbances in or above accretion discs. If the observed 154 minutes indicate an innermost stable orbital period, our upper limits of black hole mass are consistent with reverberation mapping results.

(iii) The BWB chromatic trend is dominant for 3C 273 and appears at different flux levels for intraday timescales. The BWB trend exists for short-term timescales and intermediate-term timescales but different timescales have different
correlations. There is no BWB trend for our whole time-series data sets. A significant anti-correlation between BWB
trend and length of timescales is found, which can be explained by the thermal accretion disk fluctuation model. The BWB behaviour on intraday timescales is most likely to support shock-in-jet model. The BWB trend also can be interpreted in terms of two components scenario.

(iv) By analyzing $V$-band data over 48 years, a possible quasi-periodicity of $P = 3918\pm1112$ days is found.

(v) No significant time lag between $V$ and $I$ bands is found on intraday timescales.

\begin{acknowledgements}
We sincerely thank the referee for valuable comments and suggestions. DRX thanks Chuyuan Yang, Yonggang Zheng, Hongtao Liu, Longhua Qin for useful discussions and Xuliang Fan, Shaokun Li, Liang Chen, Nenghui Liao, Jin Zhang for observations. We acknowledge the support of the staff of the
Lijiang 2.4m and Kunming 1m telescopes. Funding for the two telescopes has been provided by the Chinese Academy of Sciences and the Peoples Government of Yunnan Province. This work is financially supported by the National Nature Science Foundation of China (11663009, 11133006, 11673057, 11361140347 and U1531245), the Key Research Program of the Chinese Academy of Sciences (grant No. KJZD-EW-M06), the Strategic Priority Research Program ``The
emergence of Cosmological Structures'' of the Chinese Academy of
Sciences (grant No. XDB09000000) and the Key Laboratory of Yunnan Province University for Research in Astrophysics. DRX acknowledges support from Chinese Western Young Scholars Program and `Light of West China' Program provided by CAS. MFG acknowledges the supports from the National Science Foundation of China (grant 11473054 and U1531245) and by the Science and Technology Commission of Shanghai Municipality (14ZR1447100). This research has made use of the
NASA/IPAC Extragalactic Database (NED), that is operated by Jet
Propulsion Laboratory, California Institute of Technology, under
contract with the National Aeronautics and Space Administration.
\end{acknowledgements}

\begin{figure}
\begin{center}
\includegraphics[width=18cm,height=20cm]{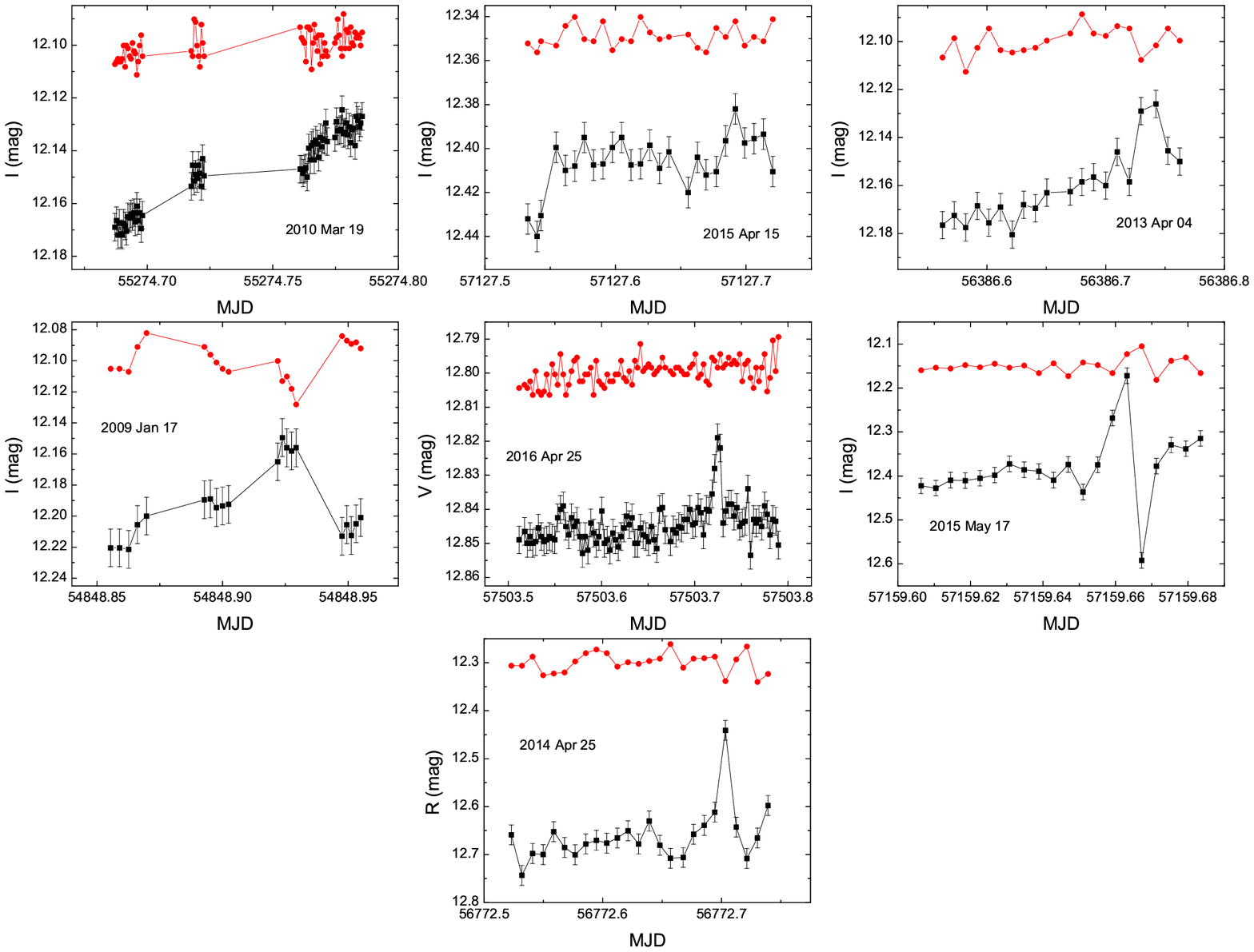}
\caption{Light curves of intraday variability for 3C 273. The black
squares and lines are the light curves for 3C 273. The red
circles and lines are the variations of $S_I$, $S_R$ and $S_V$. The
light curves of $S_I$, $S_R$ and $S_V$ are offset to avoid their
eclipsing with light curves of 3C 273. \label{fig3}}
\end{center}
\end{figure}

\begin{figure}
\begin{center}
\includegraphics[width=14cm,height=14cm]{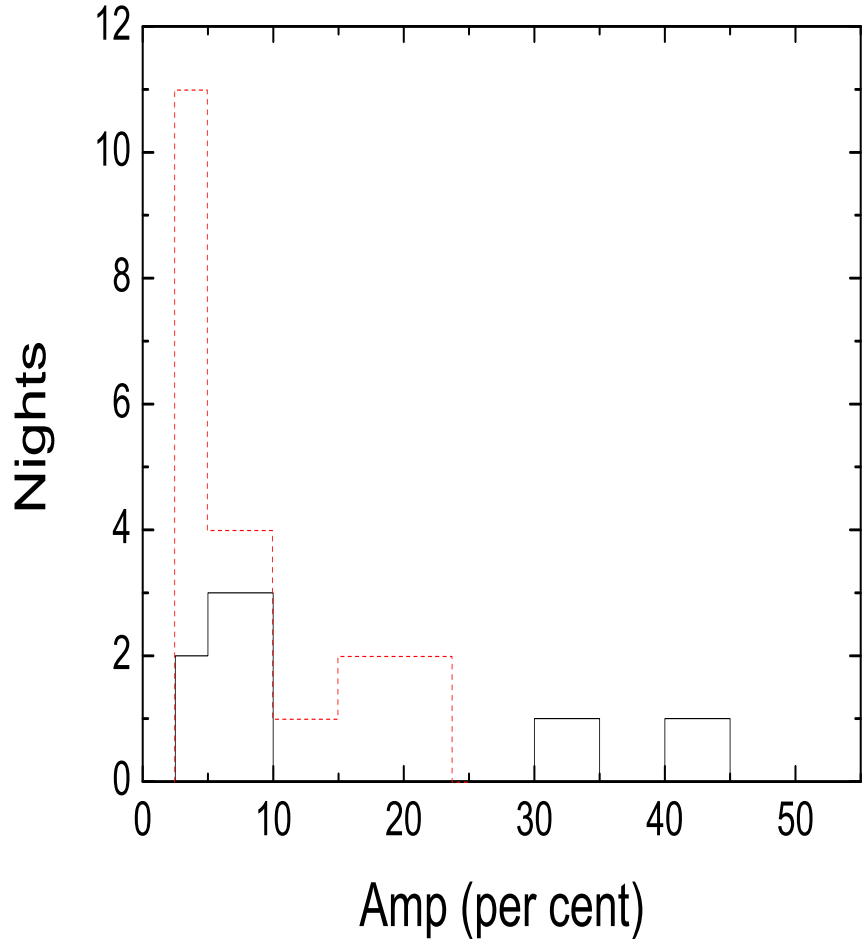}
\caption{The intraday variability amplitude distributions. The black solid lines stand for nights detected as V case, the red dashed lines for nights detected as PV case.}
\end{center}
\end{figure}

\begin{figure}
\begin{center}
\includegraphics[width=14cm,height=14cm]{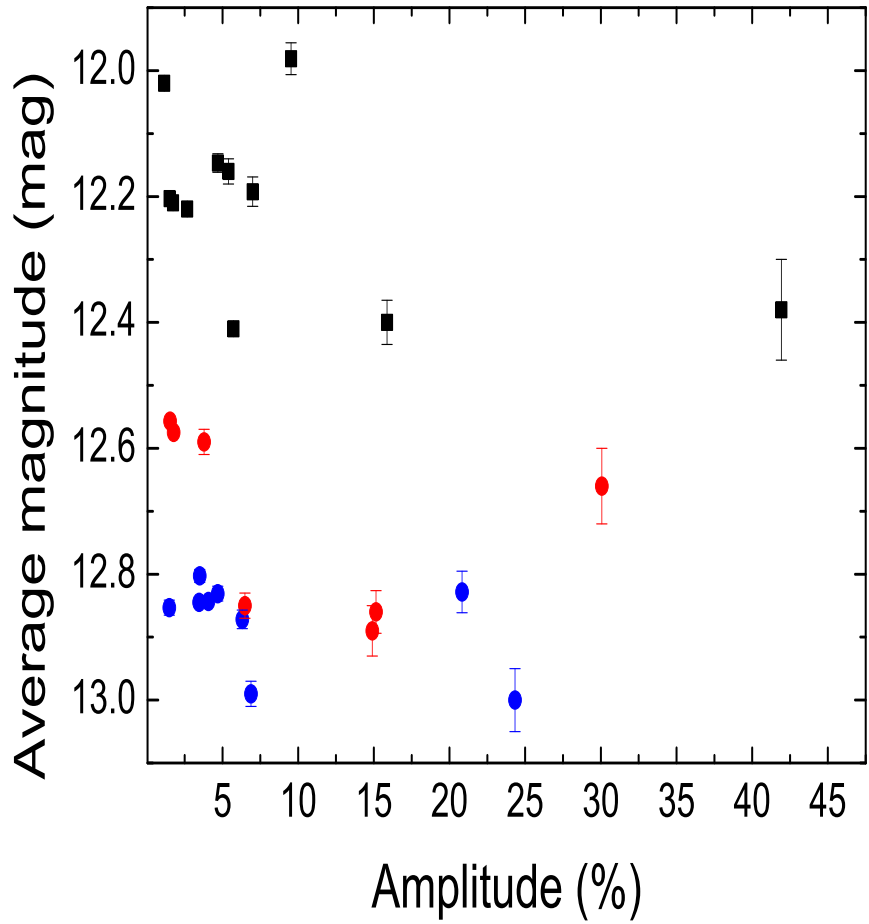}
\caption{The variability amplitude versus the average brightness. The blue circles, red circles and black
squares stand for $V$, $R$ and $I$ bands respectively.}
\end{center}
\end{figure}

\begin{figure}
\begin{center}
\includegraphics[width=17cm,height=18cm]{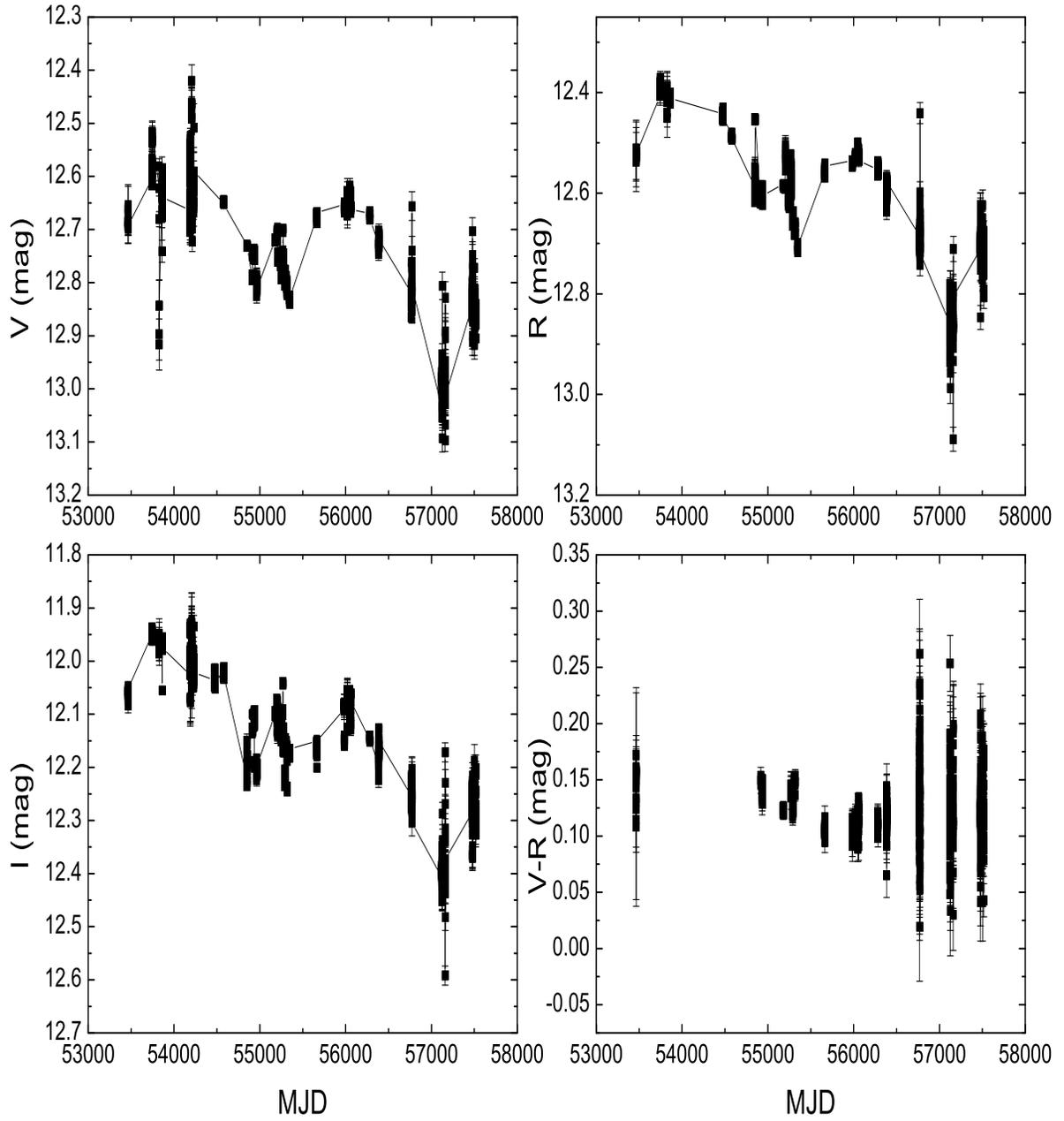}
\caption{Long-term light curves of 3C 273 in $V$, $R$, $I$ bands and color index $V-R$. \label{fig5}}
\end{center}
\end{figure}

\begin{figure}
\begin{center}
\includegraphics[angle=0,width=0.45\textwidth]{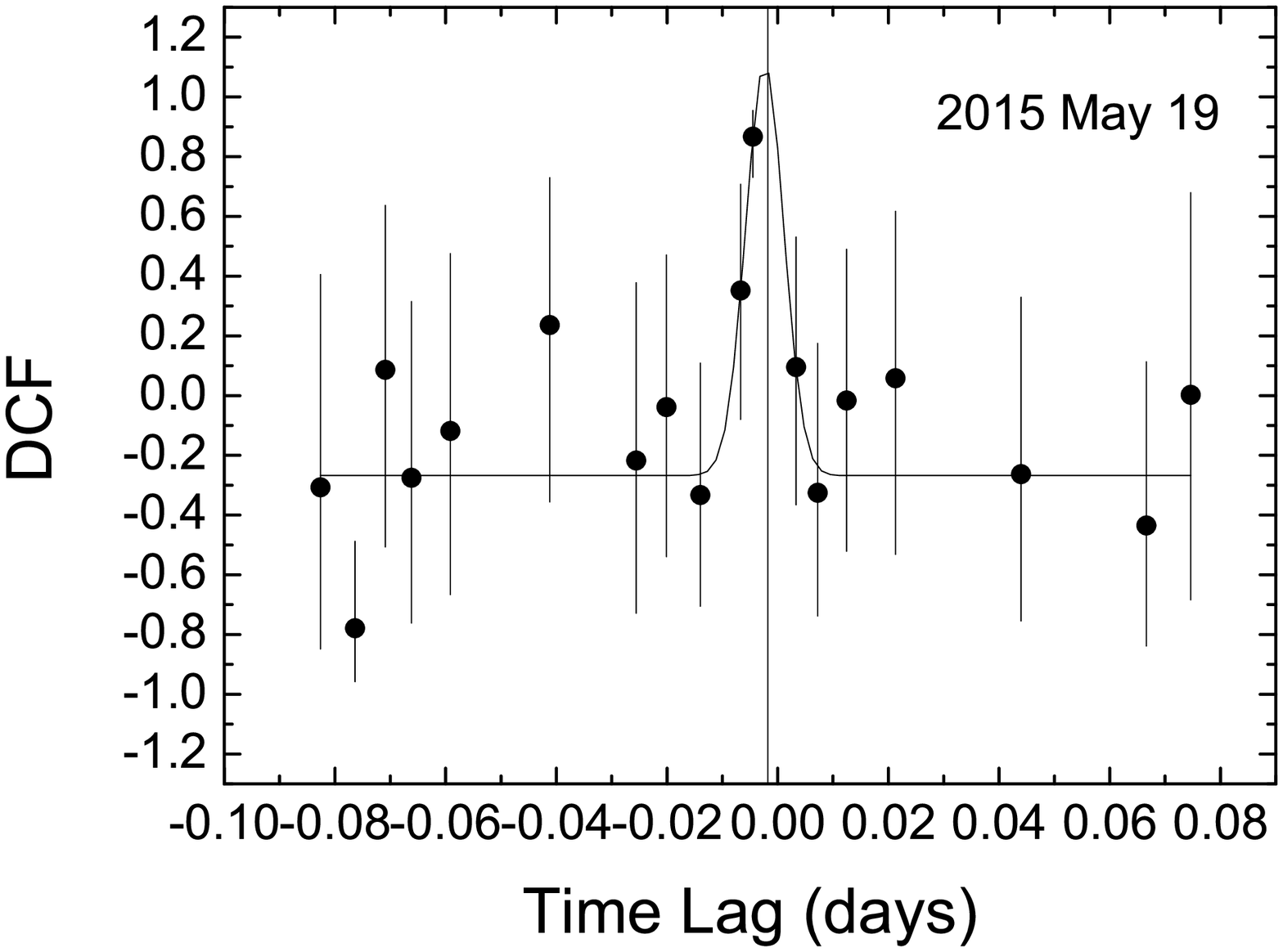}
\includegraphics[angle=0,width=0.45\textwidth]{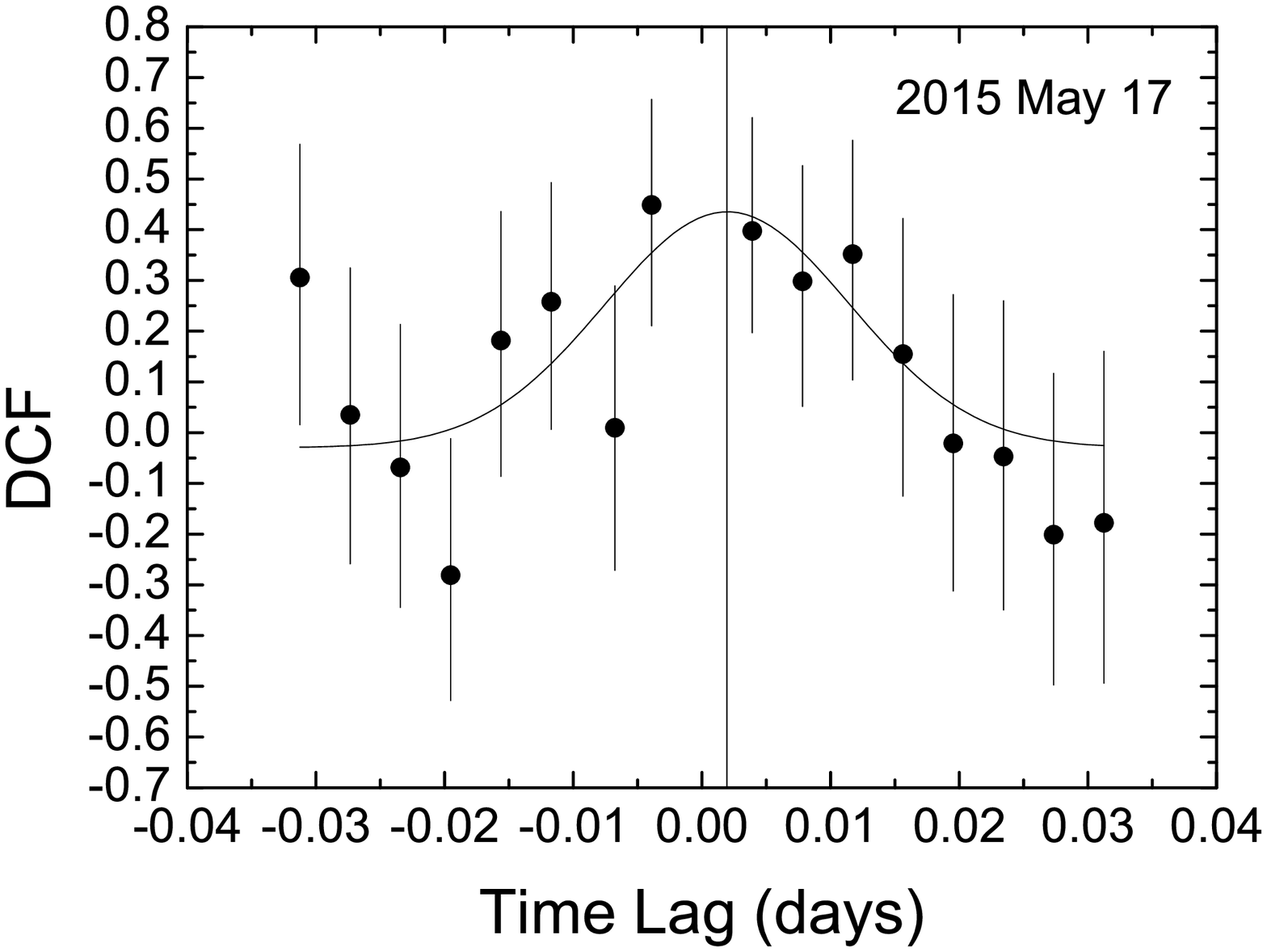}
\includegraphics[angle=0,width=0.45\textwidth]{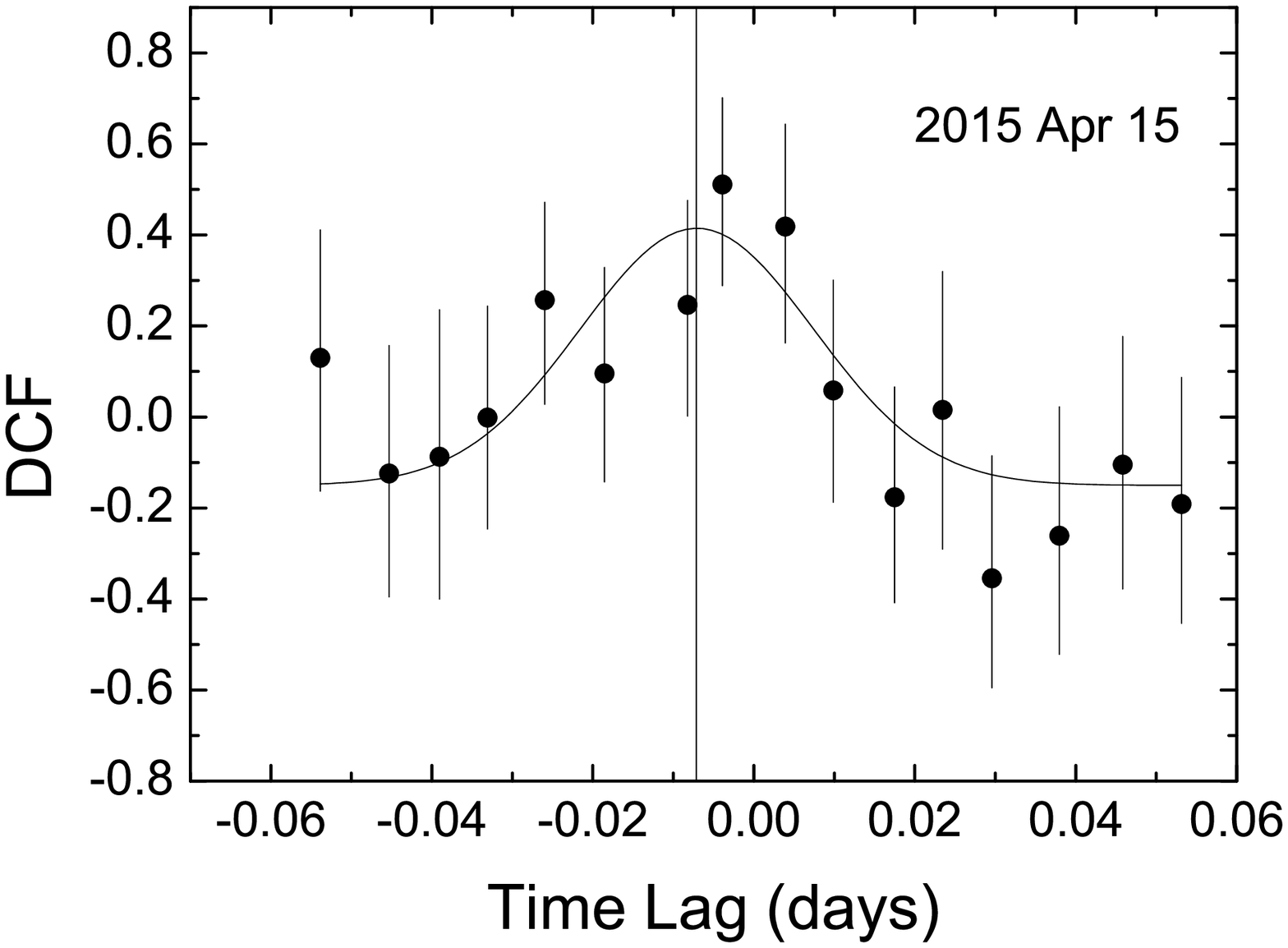}
\includegraphics[angle=0,width=0.45\textwidth]{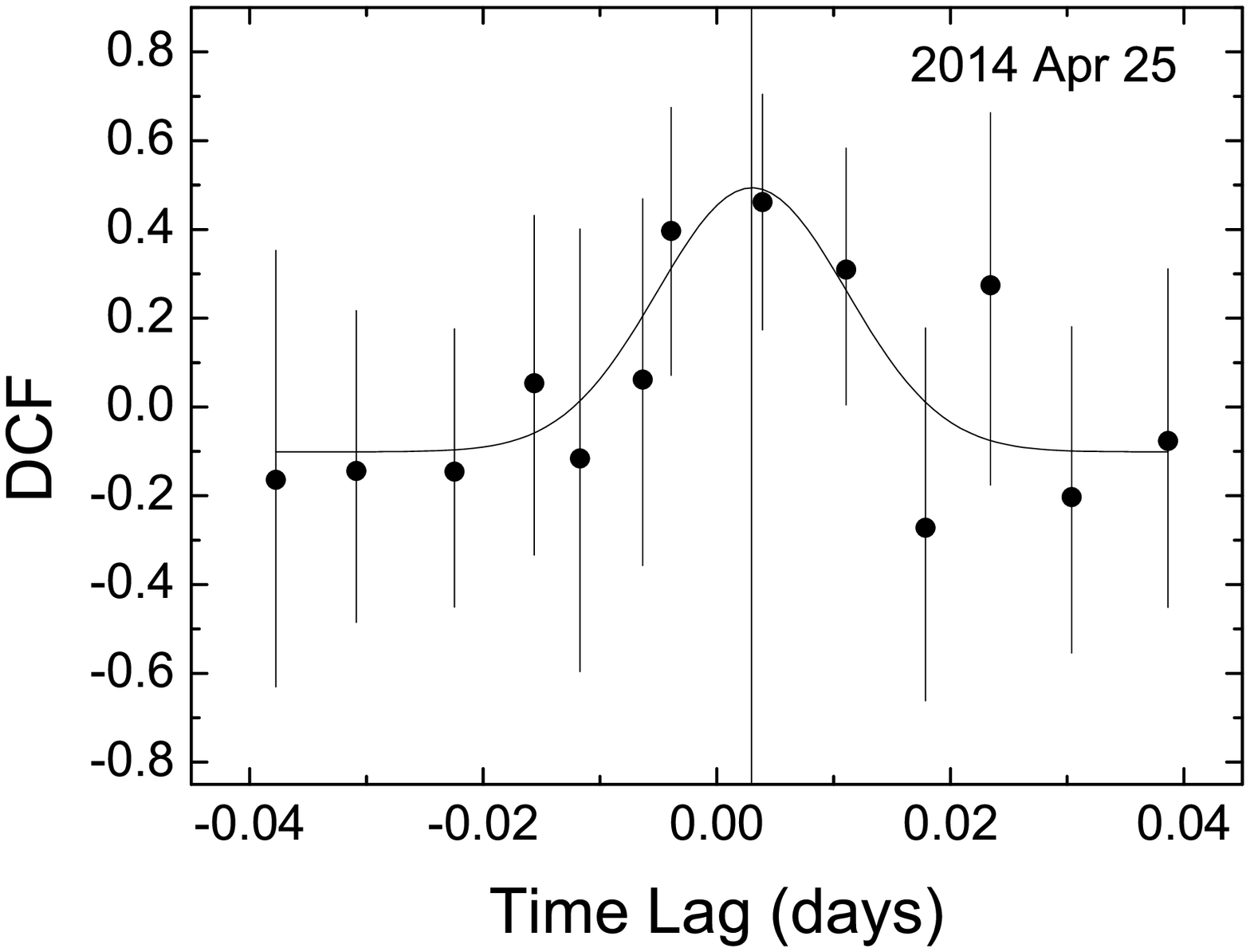}
\includegraphics[angle=0,width=0.45\textwidth]{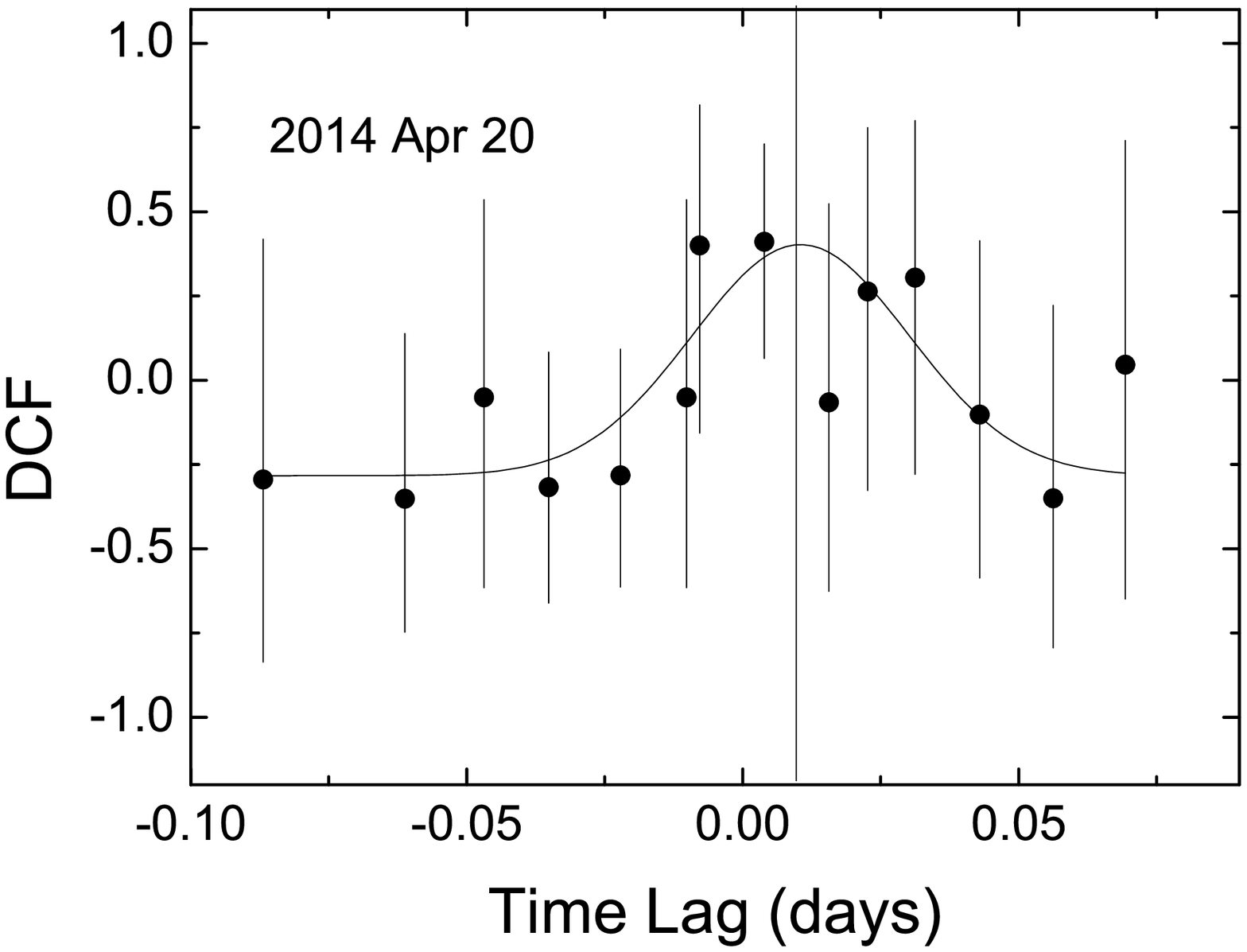}
\includegraphics[angle=0,width=0.45\textwidth]{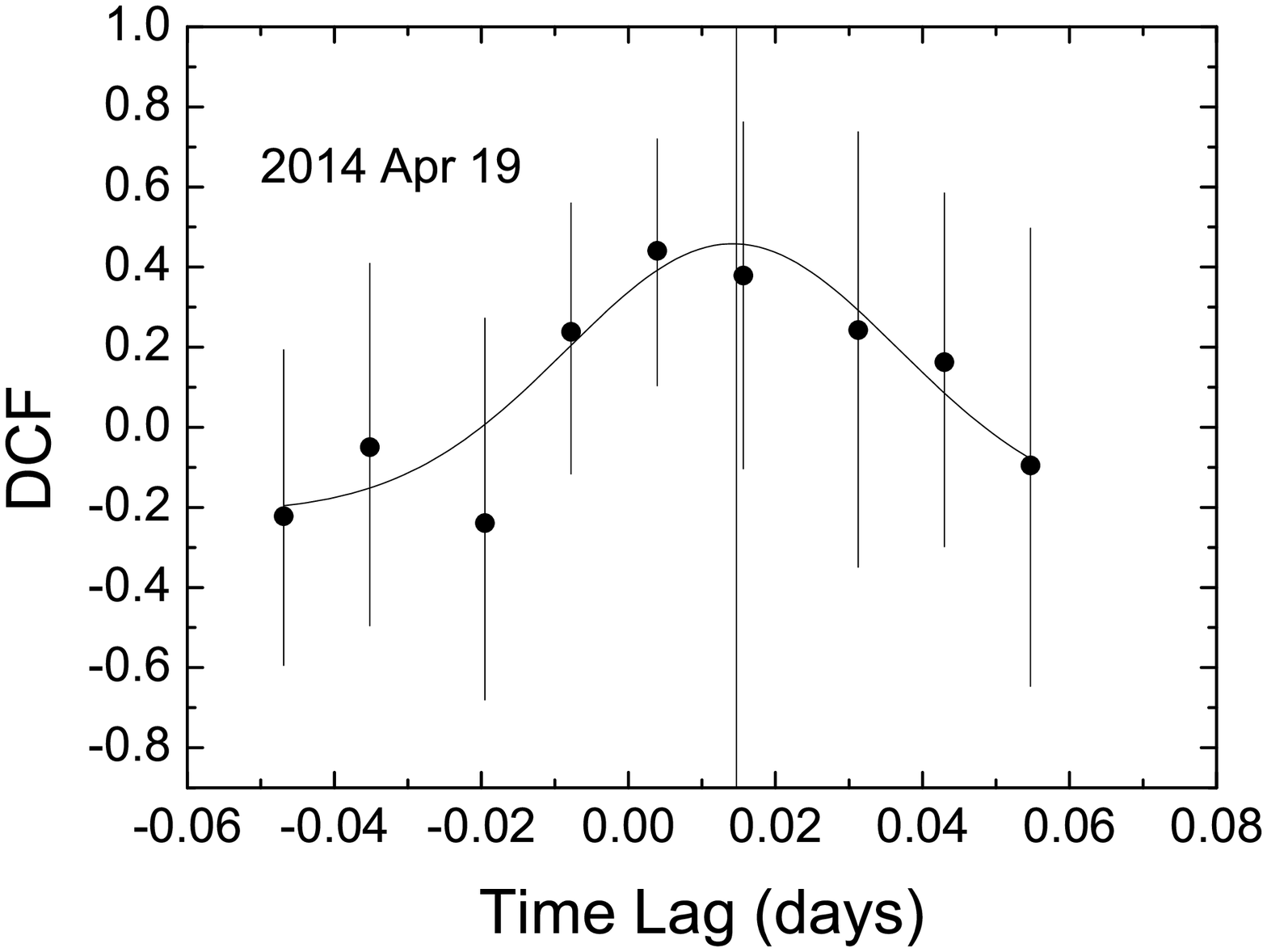}
\includegraphics[angle=0,width=0.45\textwidth]{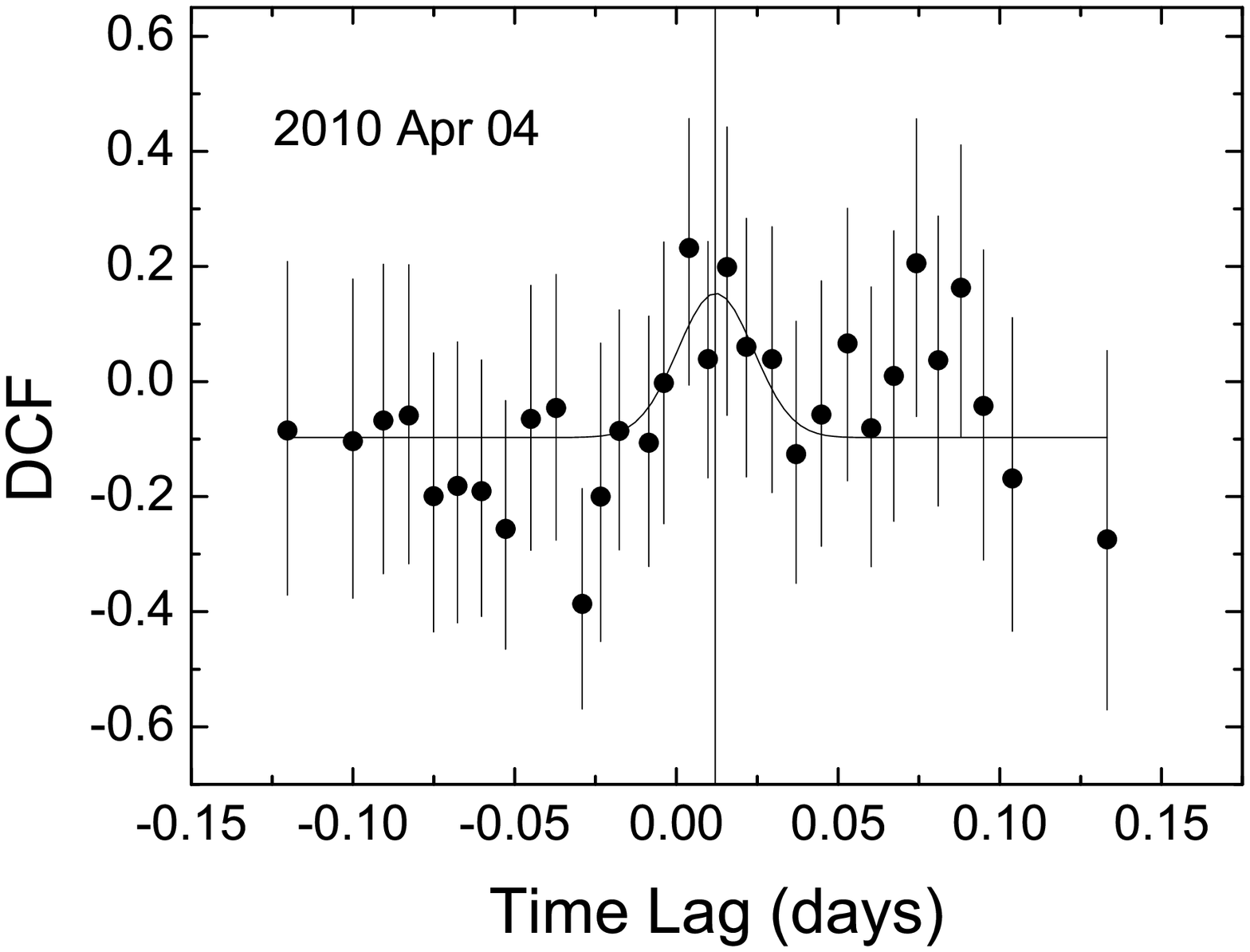}
\includegraphics[angle=0,width=0.45\textwidth]{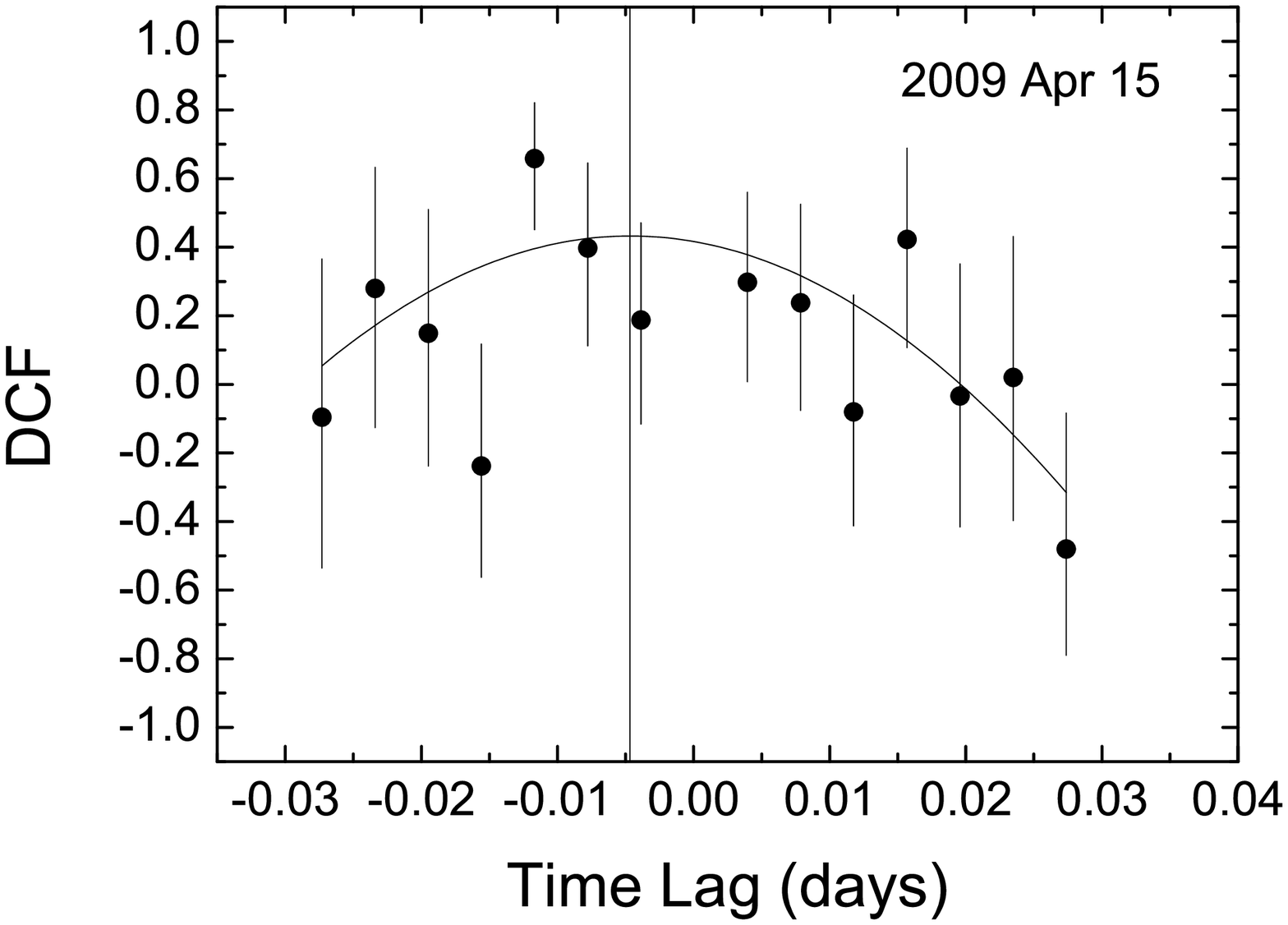}
\caption{The ZDCF plots between $V$ and $I$ bands. The curves show
Gaussian fittings to the points, and their peaks are marked with the
vertical lines.}
\end{center}
\end{figure}

\begin{figure}
\begin{center}
\includegraphics[angle=0,width=0.48\textwidth]{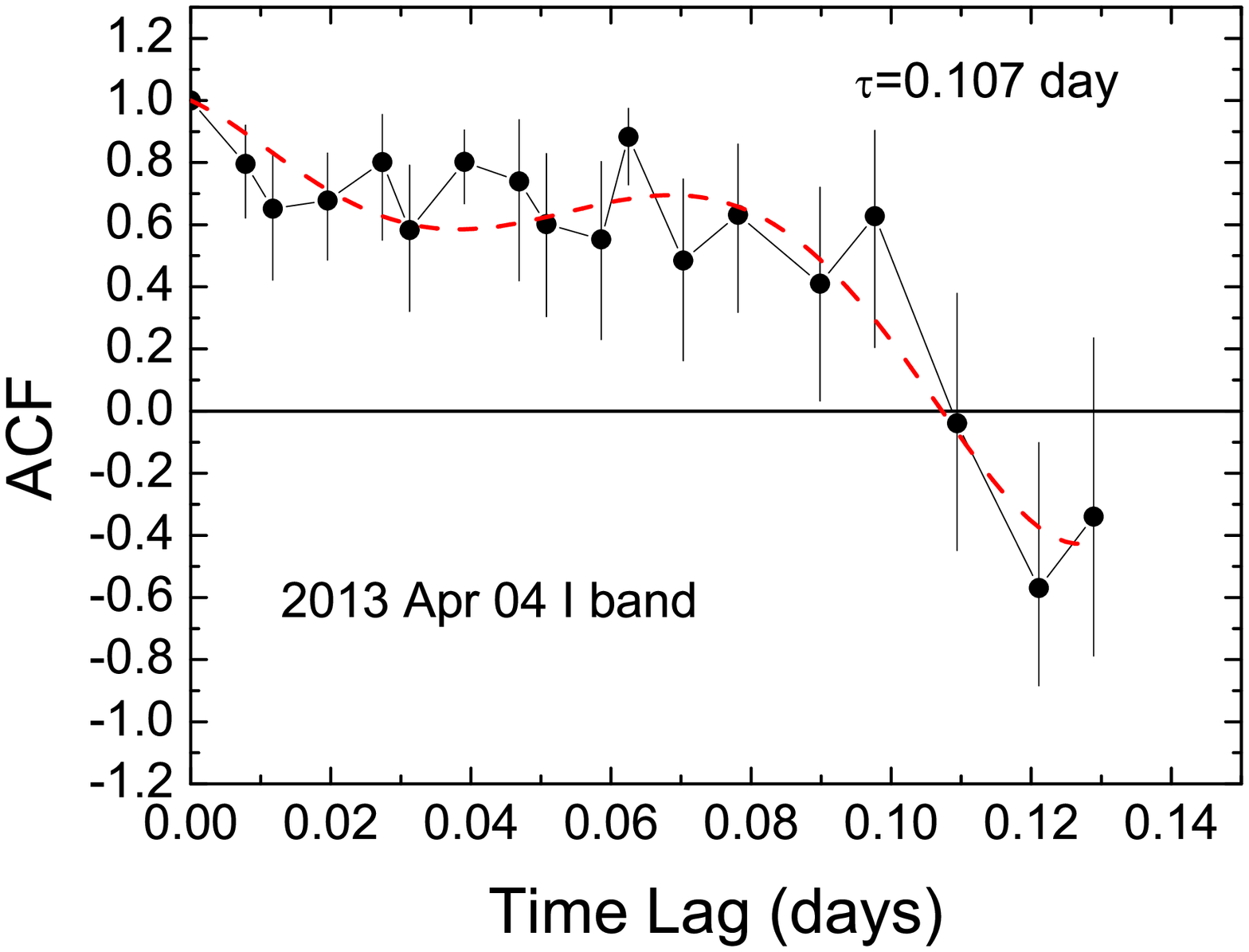}
\includegraphics[angle=0,width=0.48\textwidth]{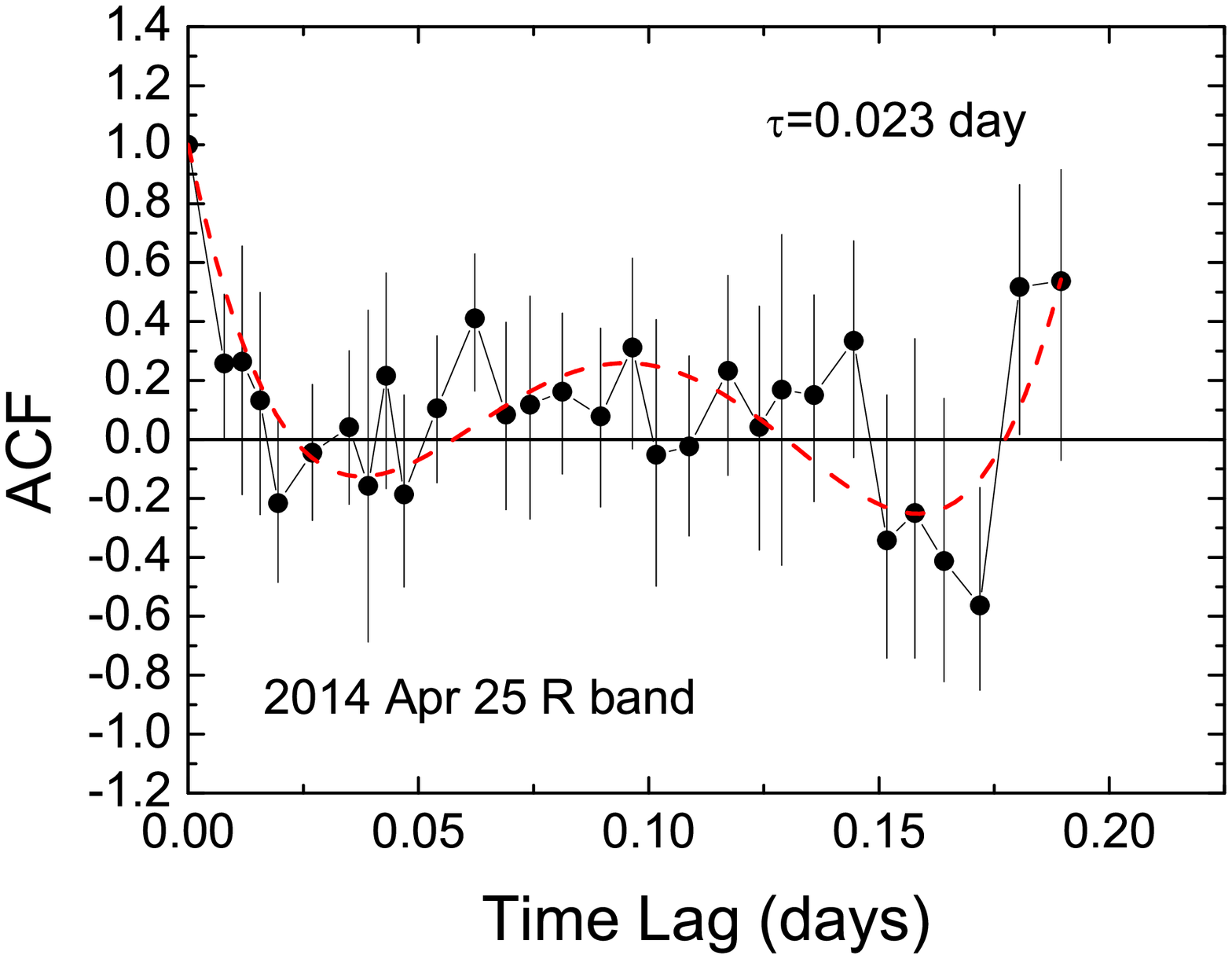}
\includegraphics[angle=0,width=0.48\textwidth]{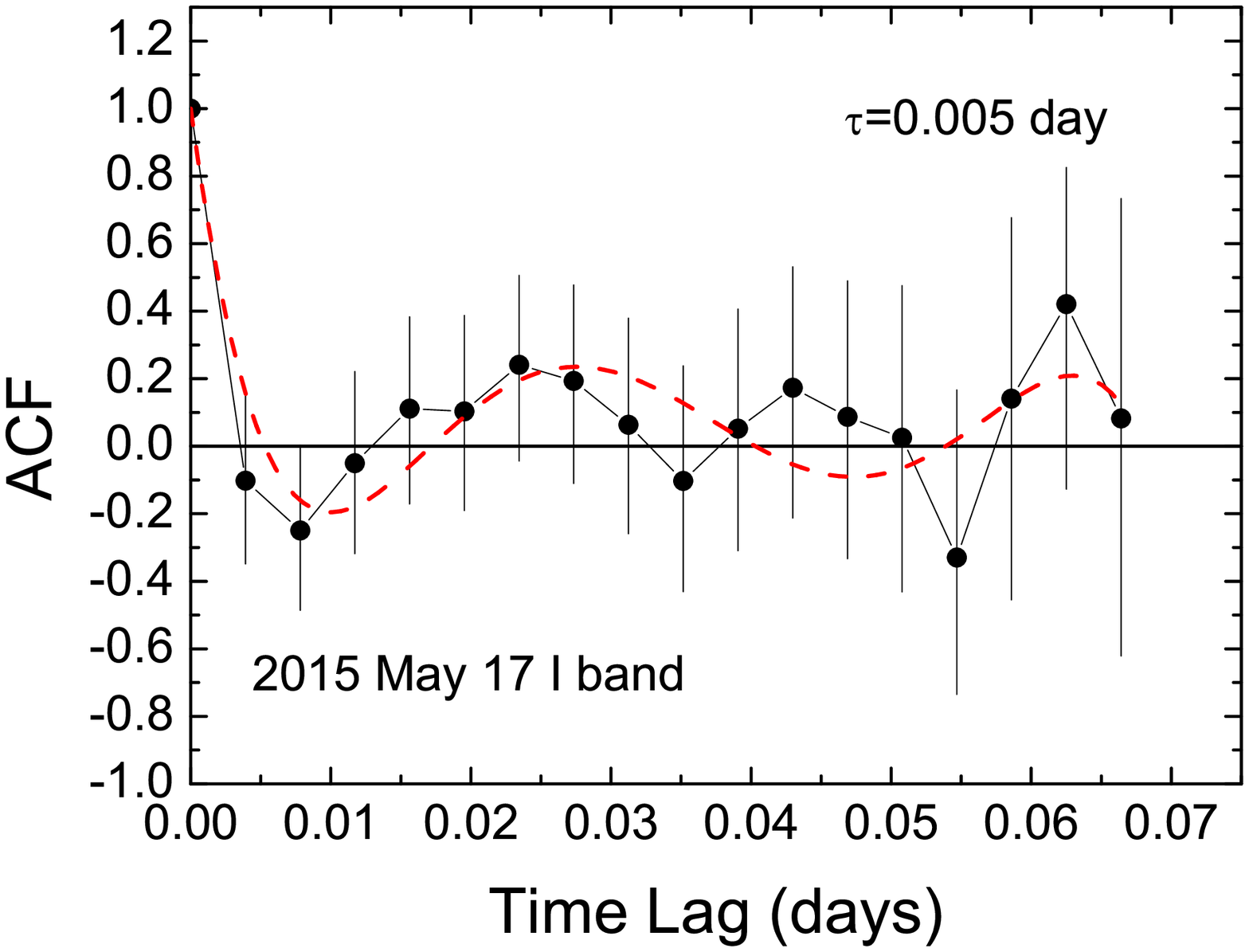}
\includegraphics[angle=0,width=0.48\textwidth]{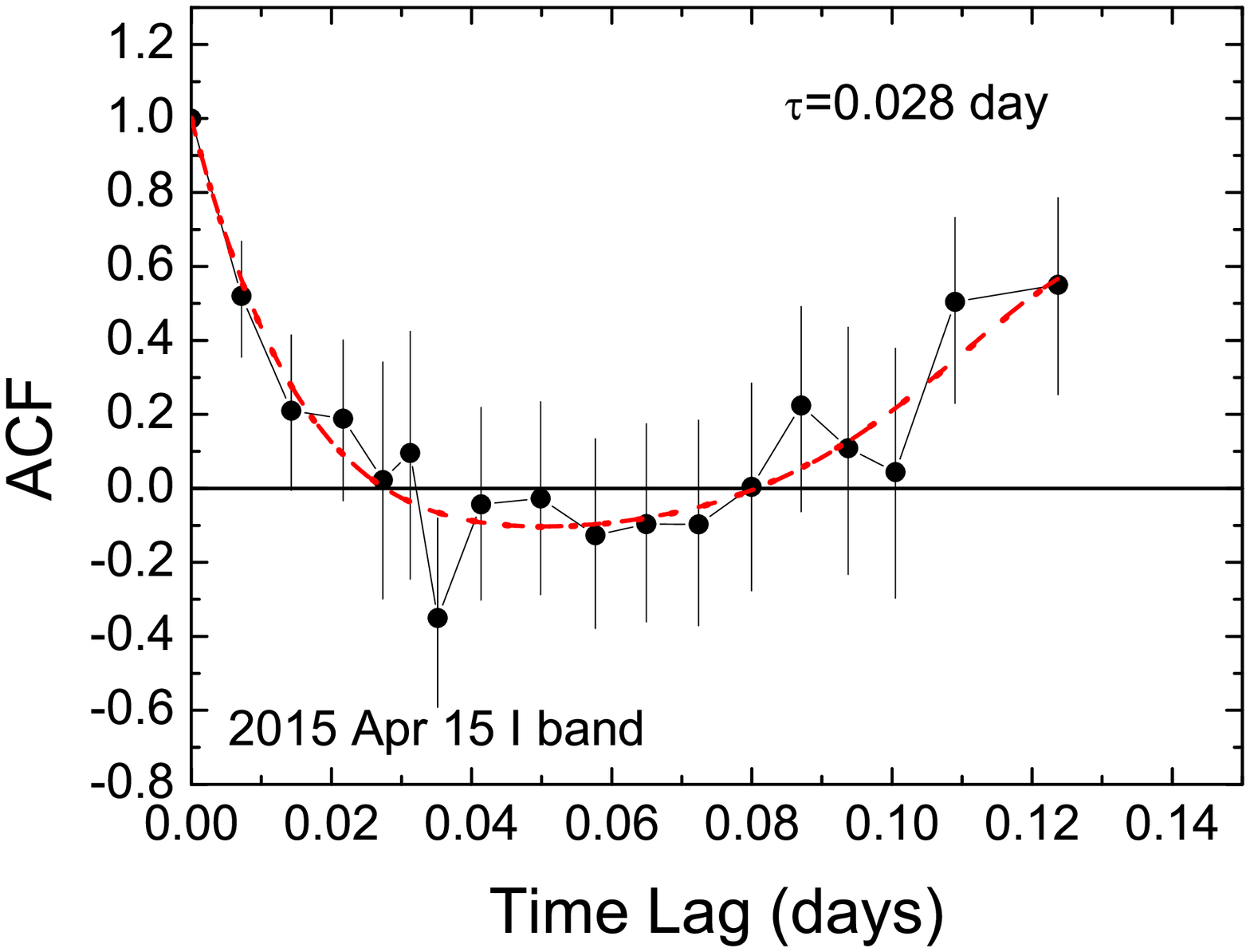}
\caption{The results of ACF analysis. The red dashed line is a fifth-order polynomial
least-squares fit.\label{fig3}}
\end{center}
\end{figure}

\begin{figure}
\begin{center}
\includegraphics[width=20cm,height=22cm]{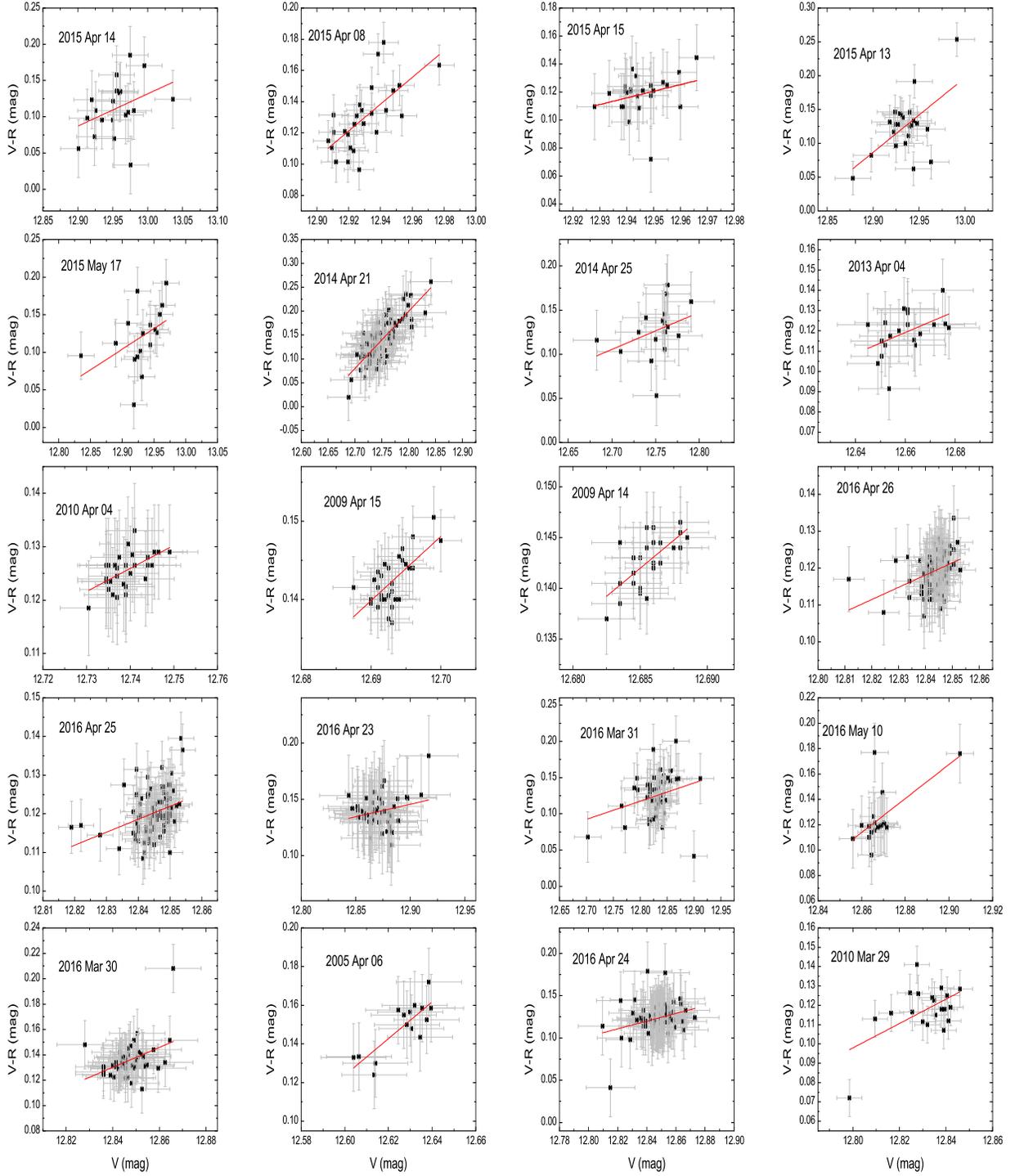}
\caption{The correlations between $V-R$ index and $V$ magnitude. The red solid lines is
results of linear regression analysis. \label{fig3}}
\end{center}
\end{figure}

\begin{figure}
\begin{center}
\includegraphics[width=16cm,height=16cm]{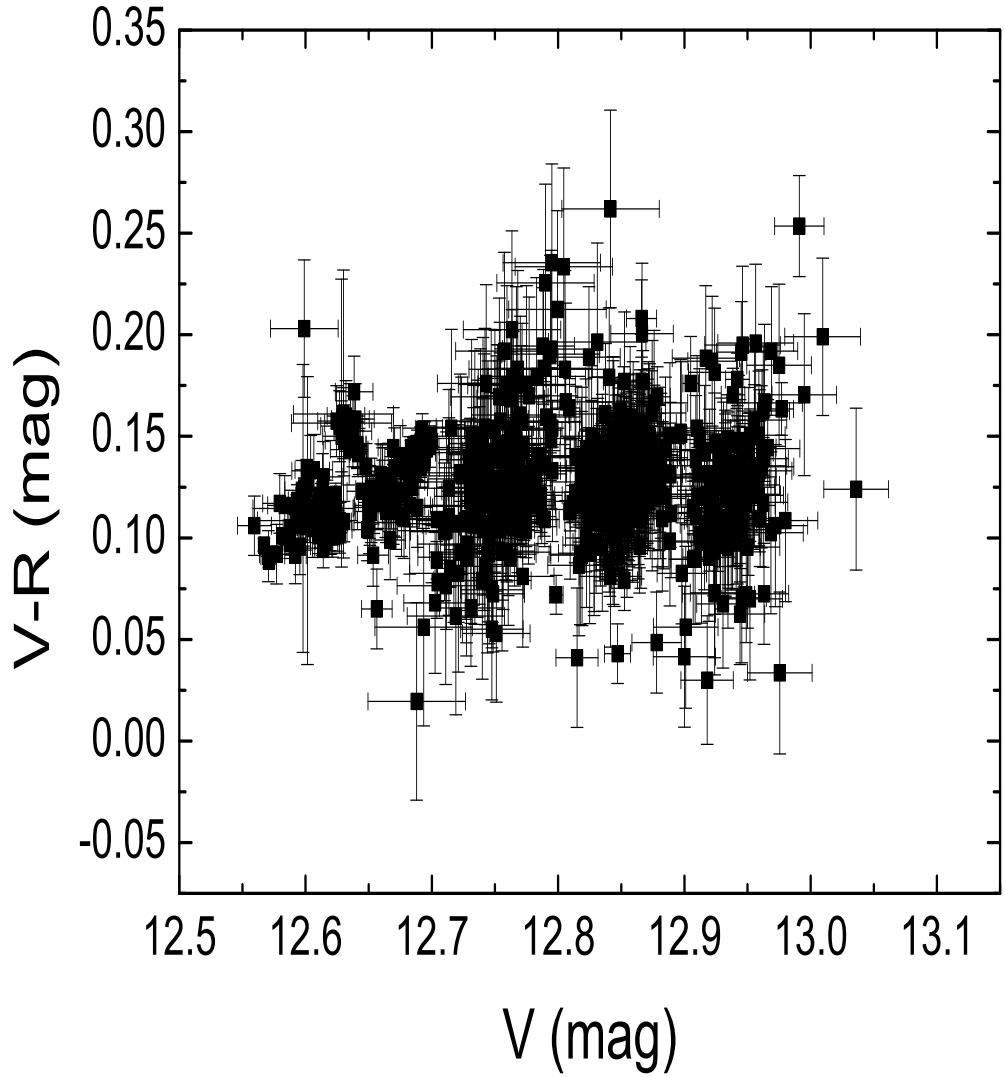}
\caption{The correlation between $V-R$ index and $V$ magnitude for
our whole time-series data sets.}
\end{center}
\end{figure}

\begin{figure}
\begin{center}
\includegraphics[width=15cm,height=15cm]{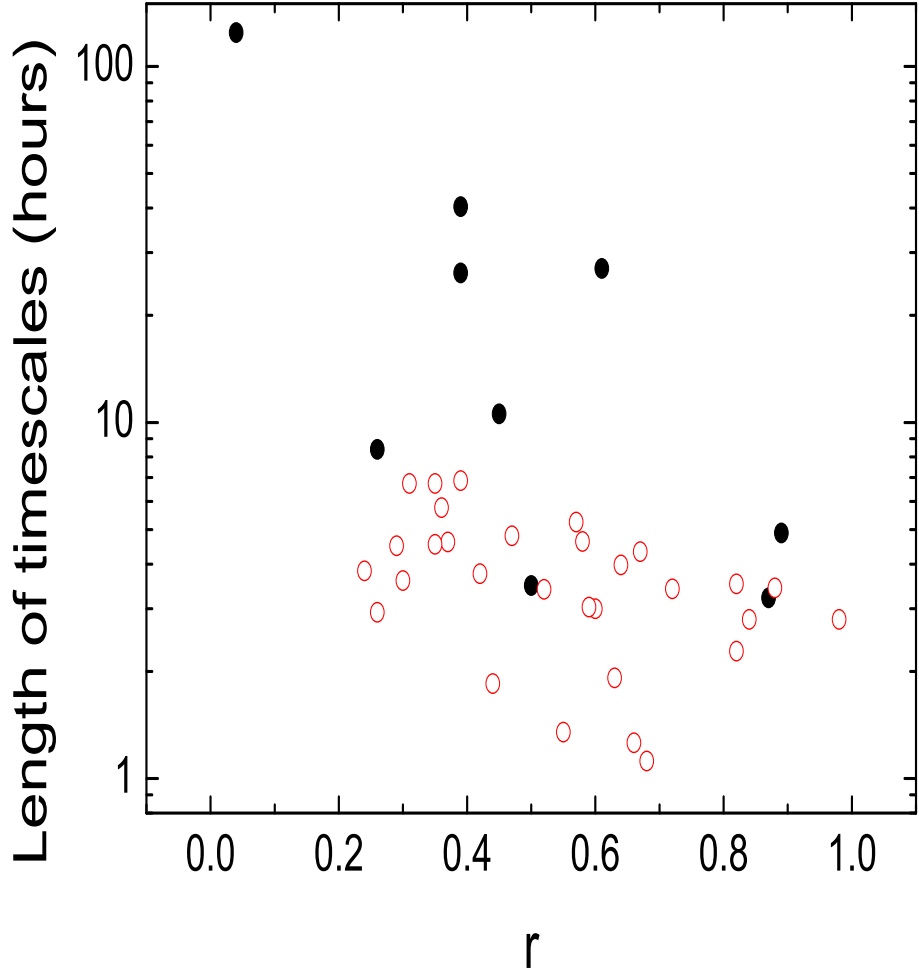}
\caption{The coefficient of correlation versus length of timescales in $V$-band. The red empty circles and black filled circles stand for intraday and longer timescales respectively.}
\end{center}
\end{figure}

\begin{figure}
\begin{center}
\includegraphics[width=15cm,height=15cm]{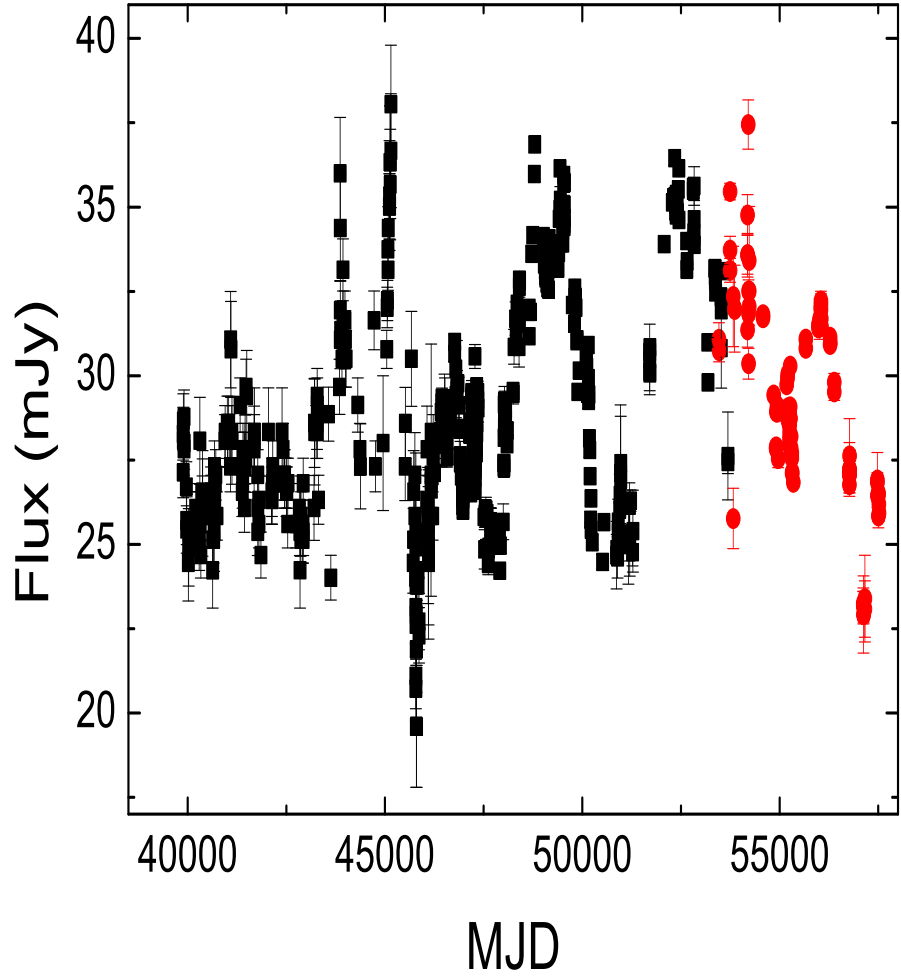}
\caption{Long-term light curves from 1968 to 2016. The red circles are our latest data and the black squares are data from Soldi et al. (2008) and Fan et al. (2014).}
\end{center}
\end{figure}

\begin{figure}
\begin{center}
\includegraphics[angle=0,width=0.5\textwidth]{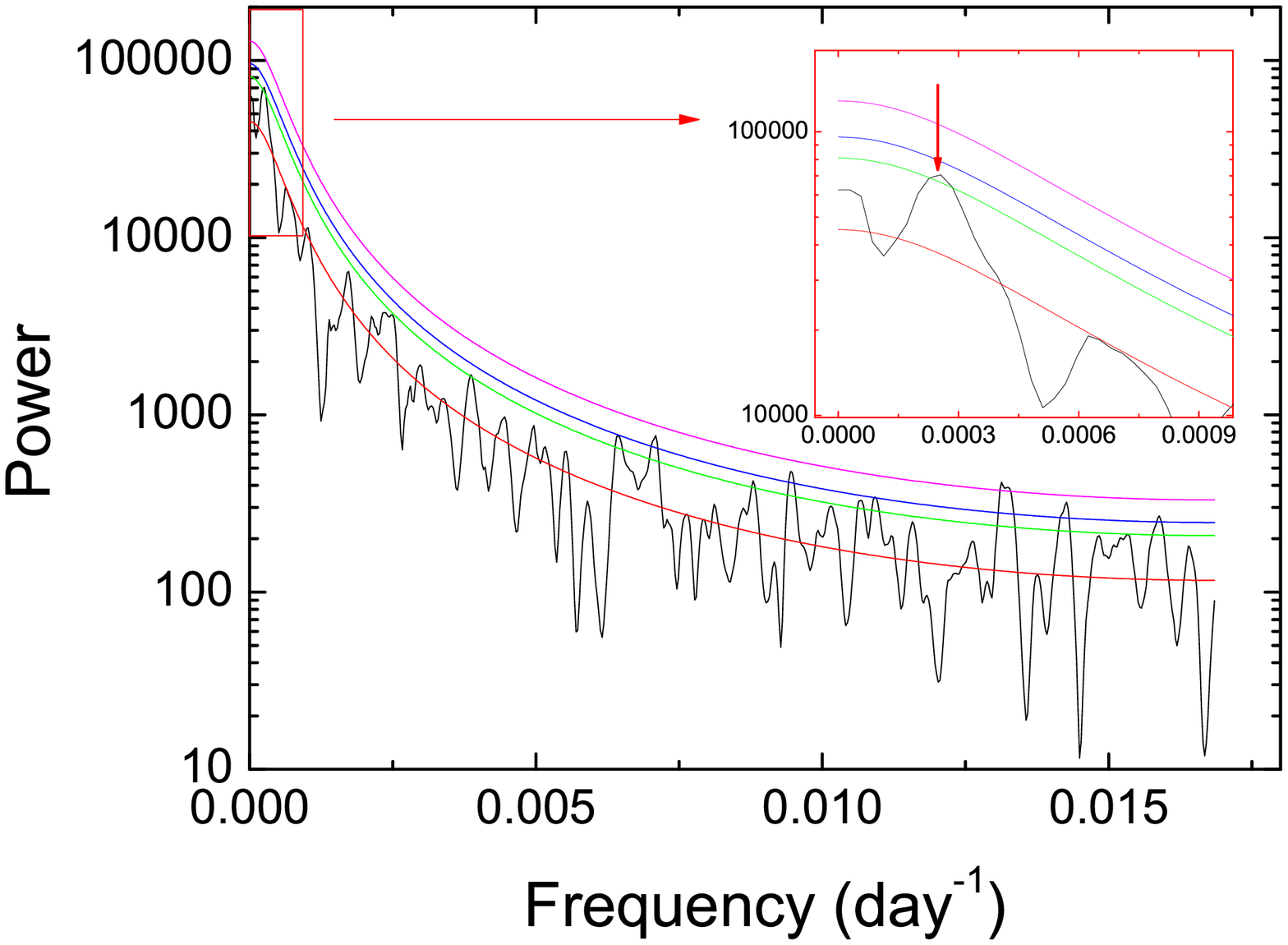}
\includegraphics[angle=0,width=0.5\textwidth]{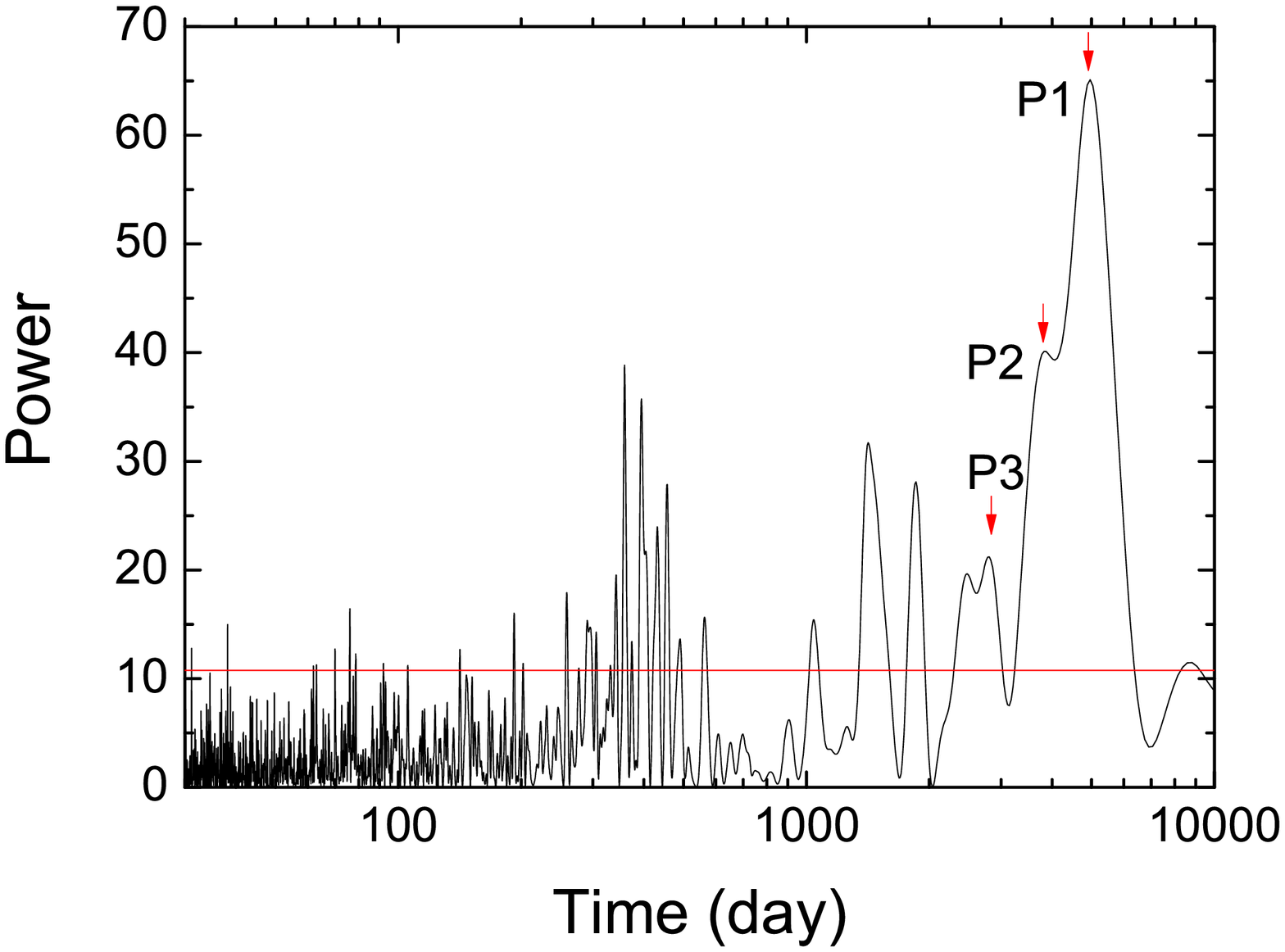}
\includegraphics[angle=0,width=0.5\textwidth]{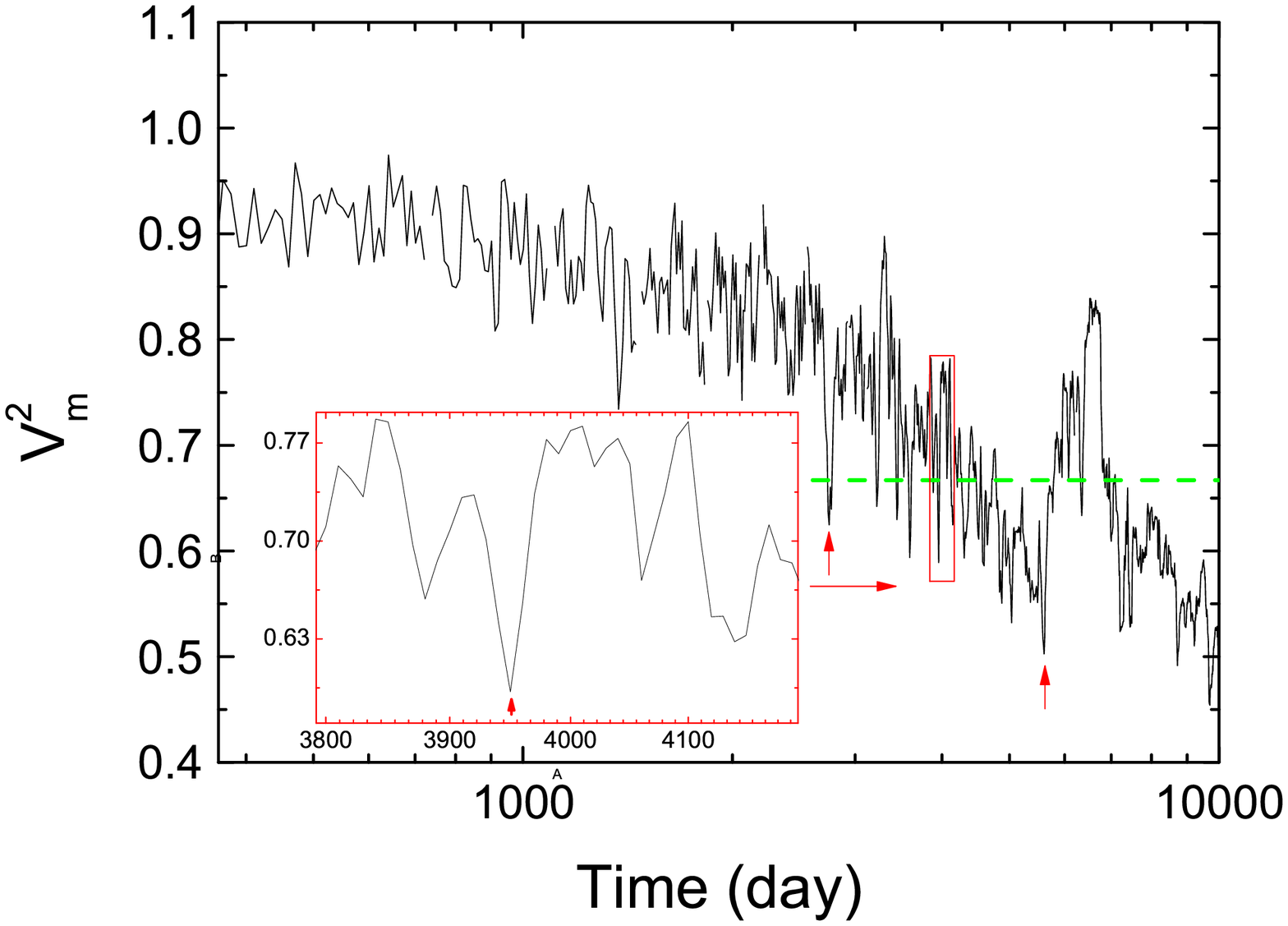}
\caption{The results of periodicity analysis. The upper, middle and bottom panels are results of REDFIT, Lomb-Scargle method and Jurkevich method respectively. For the upper panel, the black line is bias-corrected power spectra. Curves starting from the bottom are the theoretical red-noise spectrum, 90\%, 95\% and 99\% $\chi^2$ significance levels. The amplified subregion is located at upper right of the panel. The marked periodicity is $3918\pm1112$ days. For the middle panel, the red line indicates 0.01 FAP levels. The $P_1$, $P_2$ and $P_3$ peaks are $4961\pm1072$, $3848\pm471$ and $2793\pm325$ days. For the bottom panel, The amplified subregion is located at lower left of the panel. The marked periods are $2749\pm36$, $3950\pm14$ and $5594\pm44$ days. The green dashed line stands for $V_\textrm{m}^\textrm{2}=0.667$. \label{fig3}}
\end{center}
\end{figure}

\clearpage

\begin{deluxetable}{cccr}
\tablecaption{Observation log of photometric
observations\label{tbl-3}} \tablewidth{0pt}
\tablehead{\colhead{Date(UT)} & \colhead{Number(I,R,V)} &
\colhead{Time spans(h)} & \colhead{Time resolutions(min)}}
\startdata
2005 04 05	&	4,4,4	&	2.7	&	20	\\
2005 04 06	&	15,15,15	&	2.3	&	9	\\
2006 01 09$^\ast$	&	8,8,8	&	0.6	&	2	\\
2006 01 10$^\ast$	&	8,8,8	&	0.6	&	2	\\
2006 01 11$^\ast$	&	8,8,8	&	0.65	&	1.5	\\
2006 04 02$^\ast$	&	4,4,3	&	0.1	&	2	\\
2006 04 03$^\ast$	&	4,4,4	&	0.15	&	2	\\
2006 05 05$^\ast$	&	11,6,11	&	1.42,0.26,1.42	&	2.5,3,2.4	\\
2006 05 06$^\ast$	&	6,6,6	&	0.2	&	2.2	\\
2007 03 27$^\ast$	&	39,0,40	&	1,1.75	&	1.1,1.6	\\
2007 03 28$^\ast$	&	20,0,11	&	2.5,2.3	&	2,4	\\
2007 03 29$^\ast$	&	15,0,15	&	2.2	&	2.1	\\
2007 03 30$^\ast$	&	15,0,15	&	1.5	&	1.5	\\
2007 04 15$^\ast$	&	40,0,10	&	2.89,0.58	&	2.1,3.8	\\
2007 04 22$^\ast$	&	16,0,5	&	0.68,0.17	&	2.3,2.3	\\
2007 04 23$^\ast$	&	10,0,7	&	1.85,0.5	&	2.1,3.8	\\
2007 04 24$^\ast$	&	8,0,6	&	3,0.33	&	2.2,3.4	\\
2007 05 08$^\ast$	&	20,0,10	&	1.6,1.39	&	2.3,3.9	\\
2007 05 09$^\ast$	&	11,0,5	&	1.3,0.3	&	2.3,3.9	\\
2007 05 10$^\ast$	&	40,0,5	&	1.5,1.1	&	2.1,3.9	\\
2008 01 10$^\ast$	&	15,15,0	&	0.5	&	2	\\
2008 01 11$^\ast$	&	30,10,0	&	0.78,0.31	&	1.5,2	\\
2008 01 12$^\ast$	&	20,20,0	&	0.6,0.63	&	1.4,1.9	\\
2008 04 22$^\ast$	&	10,10,10	&	0.24	&	0.75	\\
2008 04 23$^\ast$	&	15,15,15	&	0.5	&	0.42	\\
2009 01 17$^\ast$	&	20,15,0	&	2.39,1.64	&	2.64,3.48	\\
2009 01 18$^\ast$	&	10,10,0	&	0.8	&	2.6	\\
2009 01 19$^\ast$	&	7,7,7	&	0.12,0.12,0.12	&	0.72,0.73,0.84	\\
2009 03 21$^\ast$	&	5,10,5	&	0.04,0.33,0.04	&	0.6,0.6,0.6	\\
2009 03 22$^\ast$	&	5,10,5	&	0.04,0.31,0.04	&	0.6,0.6,0.6	\\
2009 03 23	&	5,5,5	&	0.21	&	3.08	\\
2009 03 26	&	5,5,5	&	0.15	&	2.3	\\
2009 04 13	&	12,12,12	&	0.47	&	2.4	\\
2009 04 14	&	24,24,24	&	1.12	&	2.7	\\
2009 04 15	&	32,32,32	&	1.26	&	2.4	\\
2009 04 16$^\ast$	&	10,10,0	&	0.27	&	1.8	\\
2009 05 11$^\ast$	&	75,0,20	&	2.41,1.17	&	1.15,2.15	\\
2009 12 14	&	11,11,11	&	0.22	&	1.3	\\
2010 01 04$^\ast$	&	20,20,20	&	0.1	&	0.3	\\
2010 01 09$^\ast$	&	15,15,15	&	1	&	2.19	\\
2010 01 10$^\ast$	&	5,5,5	&	0.2	&	2	\\
2010 01 14$^\ast$	&	5,5,5	&	0.2	&	2	\\
2010 01 16$^\ast$	&	0,109,0	&	2.14	&	0.2	\\
2010 02 15$^\ast$	&	10,18,10	&	0.07,0.45,0.08	&	0.5,0.5,0.6	\\
2010 02 20$^\ast$	&	5,5,5	&	0.2	&	2	\\
2010 02 21$^\ast$	&	5,5,5	&	0.1	&	1	\\
2010 02 22$^\ast$	&	5,5,5	&	0.1	&	1	\\
2010 02 26$^\ast$	&	40,40,40	&	0.62	&	0.2	\\
2010 02 27$^\ast$	&	20,30,30	&	0.28,1.72,1.75	&	0.2,0.2,0.4	\\
2010 02 28$^\ast$	&	10,10,10	&	0.03	&	0.2	\\
2010 03 11$^\ast$	&	20,20,20	&	2.31	&	0.3	\\
2010 03 14	&	5,6,5	&	0.18	&	3	\\
2010 03 19$^\ast$	&	70,20,15	&	2.36,0.76,0.98	&	0.8,1.2,2.2	\\
2010 04 04	&	30,30,30	&	5.25	&	10	\\
2010 04 05	&	30,30,30	&	5.04	&	10	\\
2010 04 07$^\ast$	&	9,10,10	&	0.08	&	0.5	\\
2010 04 23$^\ast$	&	30,30,30	&	0.3	&	0.2	\\
2010 04 25$^\ast$	&	30,30,30	&	0.33	&	0.2	\\
2010 05 02$^\ast$	&	5,5,5	&	0.08	&	1	\\
2010 05 03$^\ast$	&	5,5,5	&	0.03,0.12,0.12	&	0.5,1.9,1.9	\\
2010 06 01$^\ast$	&	10,10,10	&	0.05	&	0.3	\\
2011 04 12	&	10,10,10	&	2.31	&	15	\\
2011 04 13	&	12,12,12	&	3.32	&	19	\\
2012 02 24	&	4,0,0	&	0.12	&	3	\\
2012 02 25	&	6,0,0	&	0.21	&	3	\\
2012 02 26	&	4,0,0	&	0.2	&	4	\\
2012 02 27	&	5,5,5	&	0.26	&	4	\\
2012 02 28	&	5,5,5	&	0.26	&	4	\\
2012 04 02	&	5,5,5	&	0.26	&	4	\\
2012 04 03	&	6,0,5	&	0.26	&	3	\\
2012 04 28	&	4,0,4	&	0.13	&	3	\\
2012 04 29	&	4,0,4	&	0.15	&	3	\\
2012 05 01	&	5,5,5	&	0.43	&	7	\\
2012 05 02	&	5,5,5	&	0.54	&	8	\\
2012 05 11	&	5,5,5	&	0.4	&	6	\\
2012 05 13	&	5,5,5	&	0.44	&	7	\\
2012 12 19	&	4,18,4	&	0.36,1.98,0.36	&	7,2.2,7	\\
2012 12 20	&	4,18,4	&	0.27,0.71,0.27	&	5.4,1.7,5.4	\\
2012 12 21	&	4,64,4	&	0.27,1.98,0.27	&	5.4,1.7,5.4	\\
2013 04 01	&	16,16,15	&	3.6	&	14	\\
2013 04 04	&	20,20,20	&	4.8	&	14	\\
2014 04 19	&	13,13,13	&	4.02	&	19	\\
2014 04 20	&	15,13,10	&	4.36	&	19	\\
2014 04 21	&	80,80,80	&	3.52	&	3	\\
2014 04 22	&	13,11,12	&	3.74	&	19	\\
2014 04 23	&	16,16,12	&	3.21	&	13	\\
2014 04 24	&	16,16,14	&	3.24	&	13	\\
2014 04 25	&	25,25,18	&	5.19	&	13	\\
2014 04 26	&	13,13,14	&	2.8	&	13	\\
2015 04 08	&	26,26,26	&	4.34	&	11	\\
2015 04 12	&	17,17,18	&	2.93	&	11	\\
2015 04 13	&	23,24,23	&	4.63	&	11	\\
2015 04 14	&	22,22,21	&	4.79	&	11	\\
2015 04 15	&	27,27,24	&	4.51	&	11	\\
2015 05 17	&	20,18,20	&	1.85	&	5.8	\\
2015 05 19	&	13,14,15	&	3.41	&	8	\\
2016 03 29	&	20,20,20	&	1.35	&	4.3	\\
2016 03 30	&	52,50,52	&	3.8	&	4.3	\\
2016 03 31	&	54,52,54	&	5.8	&	4.3	\\
2016 04 23	&	54,55,50	&	4	&	4.3	\\
2016 04 24	&	93,94,92	&	6.7	&	4.3	\\
2016 04 25	&	96,96,95	&	6.8	&	4.3	\\
2016 04 26	&	97,97,96	&	6.9	&	4.3	\\
2016 05 07	&	12,12,12	&	2	&	4.9	\\
2016 05 08	&	16,16,16	&	1.2	&	4.9	\\
2016 05 09	&	31,31,31	&	3.4	&	5	\\
2016 05 10	&	16,19,18	&	2	&	5.9	\\
2016 05 27	&	37,37,37	&	1.9	&	3	\\
\enddata
\tablecomments{The ``$^\ast$'' stands for non-quasi-simultaneous measurements in $V$, $R$ and $I$ bands (observed in the first mode).
}
\end{deluxetable}

\begin{deluxetable}{cccr}
\tablecaption{Data of $I$ Band\label{tbl-2}} \tablewidth{0pt}
\tablehead{\colhead{Date(UT)} & \colhead{MJD} & \colhead{Magnitude}
& \colhead{$\sigma$}} \startdata
2005 Apr 05	&	53464.698 	&	12.053 	&	0.014 	\\
2005 Apr 05	&	53464.717 	&	12.074 	&	0.014 	\\
2005 Apr 05	&	53464.731 	&	12.084 	&	0.014 	\\
2005 Apr 05	&	53464.811 	&	12.055 	&	0.014 	\\
2005 Apr 06	&	53465.685 	&	12.057 	&	0.011 	\\
\enddata
\tablecomments{Column (1) is the universal time (UT) of observation,
column (2) the corresponding modified Julian day (MJD), column (3)
the magnitude, column (4) the rms error. Table 2 is available in its entirety in the
electronic edition of the {\sl The Astrophysical Journal
Supplement}. A portion is shown here for guidance regarding its form
and content.}
\end{deluxetable}

\begin{deluxetable}{cccr}
\tablecaption{Data of $R$ Band\label{tbl-3}} \tablewidth{0pt}
\tablehead{\colhead{Date(UT)} & \colhead{MJD} & \colhead{Magnitude}
& \colhead{$\sigma$}} \startdata
2005 Apr 05	&	53464.703 	&	12.538 	&	0.059 	\\
2005 Apr 05	&	53464.719 	&	12.514 	&	0.059 	\\
2005 Apr 05	&	53464.733 	&	12.529 	&	0.059 	\\
2005 Apr 05	&	53464.815 	&	12.517 	&	0.059 	\\
2005 Apr 06	&	53465.687 	&	12.530 	&	0.009 	\\
\enddata
\tablecomments{The meanings of columns are same with Table 2. Table
3 is available in its entirety in the electronic edition of the {\sl
The Astrophysical Journal Supplement}. A portion is shown here for
guidance regarding its form and content.}
\end{deluxetable}

\begin{deluxetable}{cccr}
\tablecaption{Data of $V$ Band\label{tbl-4}} \tablewidth{0pt}
\tablehead{\colhead{Date(UT)} & \colhead{MJD} & \colhead{Magnitude}
& \colhead{$\sigma$}} \startdata
2005 Apr 05	&	53464.709 	&	12.659 	&	0.040 	\\
2005 Apr 05	&	53464.723 	&	12.687 	&	0.040 	\\
2005 Apr 05	&	53464.737 	&	12.655 	&	0.040 	\\
2005 Apr 05	&	53464.819 	&	12.686 	&	0.040 	\\
2005 Apr 06	&	53465.688 	&	12.695 	&	0.015 	\\
\enddata
\tablecomments{The meanings of columns are same with Table 2. Table
4 is available in its entirety in the electronic edition of the {\sl
The Astrophysical Journal Supplement}. A portion is shown here for
guidance regarding its form and content.}
\end{deluxetable}

\begin{deluxetable}{ccccccccccr}
\small
\tablecaption{Results of IDV Observations of 3C 273\label{tbl-5}}
\tablewidth{0pt} \tablehead{\colhead{Date} & \colhead{Band}  &	\colhead{N}	&	T(h) & \colhead{$F$} &\colhead{$F_C(99)$}
&\colhead{$F_A$} &\colhead{$F_A(99)$} &\colhead{V/N} &\colhead{A(\%)} &\colhead{Ave(mag)}\\
\colhead{(1)} & \colhead{(2)}  & \colhead{(3)} &
\colhead{(4)} &\colhead{(5)} &\colhead{(6)} &\colhead{(7)}
&\colhead{(8)} &\colhead{(9)} &\colhead{(10)}&\colhead{(11)}} \startdata
2007 Mar 27	&	V	&	40	&	1.75	&	0.76	&	2.16	&	1.76	&	3.26	&	N	&	6.24	&	12.66	\\
2007 Mar 28	&	I	&	20	&	2.48	&	1.03	&	3.03	&	4.24	&	4.89	&	N	&	1.65	&	12.00	\\
2007 Apr 15	&	I	&	40	&	2.89	&	0.45	&	2.16	&	5.67	&	3.26	&	PV	&	9.53	&	11.98	\\
2007 May 08	&	I	&	20	&	1.59	&	0.68	&	3.03	&	0.66	&	4.89	&	N	&	2.71	&	12.03	\\
2007 May 10	&	I	&	40	&	1.48	&	0.52	&	2.16	&	0.96	&	3.26	&	N	&	3.03	&	12.01	\\
2008 Jan 11	&	I	&	30	&	0.78	&	0.85	&	2.42	&	4.56	&	3.90	&	PV	&	1.15	&	12.02	\\
2008 Jan 12	&	I	&	20	&	0.48	&	0.63	&	3.03	&	1.34	&	5.29	&	N	&	1.18	&	12.05	\\
2008 Jan 12	&	R	&	20	&	0.63	&	0.76	&	3.03	&	0.08	&	5.29	&	N	&	1.22	&	12.45	\\
2009 Jan 17	&	I	&	20	&	2.39	&	3.93	&	3.03	&	21.15	&	4.70	&	V	&	6.99	&	12.19	\\
2009 May 11	&	I	&	75	&	2.41	&	0.41	&	1.75	&	4.11	&	2.44	&	PV	&	1.50	&	12.20	\\
2009 May 11	&	V	&	20	&	1.17	&	0.87	&	3.03	&	10.72	&	5.29	&	PV	&	3.50	&	12.80	\\
2010 Mar 19	&	I	&	70	&	2.36	&	3.74	&	1.75	&	109.00	&	2.47	&	V	&	4.70	&	12.15	\\
2010 Mar 19	&	R	&	20	&	0.76	&	0.57	&	3.03	&	15.62	&	5.29	&	PV	&	1.77	&	12.57	\\
2010 Apr 04	&	I	&	30	&	5.21	&	0.66	&	2.42	&	5.04	&	3.46	&	PV	&	2.66	&	12.22	\\
2010 Apr 04	&	R	&	30	&	5.23	&	0.61	&	2.42	&	2.95	&	3.90	&	N	&	0.85	&	12.66	\\
2010 Apr 04	&	V	&	30	&	5.25	&	0.68	&	2.42	&	3.39	&	3.46	&	N	&	1.60	&	12.80	\\
2010 Apr 05	&	I	&	30	&	5.04	&	1.00	&	2.42	&	13.57	&	3.46	&	PV	&	1.73	&	12.21	\\
2010 Apr 05	&	R	&	30	&	5.04	&	0.93	&	2.42	&	3.00	&	3.46	&	N	&	1.83	&	12.64	\\
2010 Apr 05	&	V	&	30	&	5.04	&	0.55	&	2.42	&	2.06	&	3.46	&	N	&	0.59	&	12.79	\\
2012 Dec 21	&	R	&	64	&	1.98	&	0.97	&	1.82	&	4.69	&	2.35	&	PV	&	1.54	&	12.56	\\
2013 Apr 04	&	I	&	20	&	4.80	&	7.35	&	3.03	&	9.03	&	4.70	&	V	&	5.39	&	12.16	\\
2013 Apr 04	&	R	&	20	&	4.80	&	1.53	&	3.03	&	4.97	&	4.70	&	PV	&	3.78	&	12.59	\\
2013 Apr 04	&	V	&	20	&	4.81	&	0.86	&	3.03	&	2.34	&	4.70	&	N	&	2.75	&	12.72	\\
2014 Apr 21	&	V	&	80	&	3.52	&	0.94	&	1.69	&	0.75	&	2.13	&	N	&	20.23	&	12.81	\\
2014 Apr 25	&	I	&	25	&	5.19	&	1.07	&	2.66	&	3.02	&	3.93	&	N	&	6.91	&	12.24	\\
2014 Apr 25	&	R	&	25	&	5.19	&	7.65	&	2.66	&	2.42	&	3.93	&	V	&	30.06	&	12.66	\\
2015 Apr 08	&	I	&	26	&	4.34	&	1.06	&	2.60	&	1.27	&	3.84	&	N	&	4.39	&	12.40	\\
2015 Apr 08	&	R	&	26	&	4.34	&	3.26	&	2.60	&	1.76	&	3.84	&	PV	&	6.48	&	12.85	\\
2015 Apr 08	&	V	&	26	&	4.34	&	3.33	&	2.60	&	0.22	&	3.84	&	PV	&	6.88	&	12.99	\\
2015 Apr 13	&	I	&	23	&	4.58	&	2.57	&	2.79	&	1.08	&	4.20	&	N	&	9.04	&	12.40	\\
2015 Apr 13	&	R	&	24	&	4.60	&	5.15	&	2.79	&	0.53	&	4.02	&	PV	&	15.14	&	12.86	\\
2015 Apr 13	&	V	&	23	&	4.63	&	1.51	&	2.78	&	1.47	&	4.20	&	N	&	8.04	&	12.99	\\
2015 Apr 14	&	I	&	22	&	4.79	&	2.93	&	2.86	&	1.25	&	4.30	&	PV	&	15.87	&	12.40	\\
2015 Apr 14	&	R	&	20	&	4.62	&	3.90	&	2.86	&	0.13	&	4.30	&	PV	&	14.90	&	12.89	\\
2015 Apr 14	&	V	&	20	&	4.62	&	4.74	&	2.94	&	1.43	&	4.46	&	PV	&	24.33	&	13.00	\\
2015 Apr 15	&	I	&	27	&	4.51	&	3.58	&	2.55	&	8.30	&	3.70	&	V	&	5.72	&	12.41	\\
2015 Apr 15	&	R	&	26	&	4.51	&	0.63	&	2.55	&	1.88	&	3.70	&	N	&	3.16	&	12.87	\\
2015 Apr 15	&	V	&	24	&	4.51	&	2.04	&	2.72	&	3.91	&	4.03	&	N	&	3.69	&	13.00	\\
2015 May 17	&	I	&	20	&	1.85	&	20.68	&	3.03	&	2.02	&	4.62	&	V	&	41.93	&	12.38	\\
2016 Mar 29	&	V	&	20	&	1.35	&	5.06	&	3.03	&	1.80	&	4.70	&	PV	&	4.69	&	12.83	\\
2016 Mar 30	&	V	&	52	&	3.76	&	0.64	&	1.95	&	0.85	&	2.57	&	N	&	3.40	&	12.85	\\
2016 Mar 31	&	V	&	54	&	5.77	&	2.02	&	1.92	&	1.67	&	2.51	&	PV	&	20.83	&	12.83	\\
2016 Apr 23	&	V	&	50	&	3.83	&	0.56	&	1.95	&	4.91	&	2.62	&	PV	&	6.31	&	12.87	\\
2016 Apr 24	&	V	&	92	&	6.74	&	0.84	&	1.63	&	1.44	&	2.04	&	N	&	9.87	&	12.85	\\
2016 Apr 25	&	V	&	95	&	6.74	&	2.27	&	1.63	&	4.18	&	2.00	&	V	&	3.45	&	12.84	\\
2016 Apr 26	&	V	&	96	&	6.87	&	1.26	&	1.62	&	2.30	&	2.00	&	PV	&	4.07	&	12.84	\\
2016 May 09	&	V	&	31	&	3.40	&	0.79	&	2.39	&	0.28	&	3.40	&	N	&	3.51	&	12.86	\\
2016 May 27	&	V	&	37	&	1.86	&	0.45	&	2.21	&	8.85	&	3.06	&	PV	&	1.48	&	12.85	\\
\enddata
\tablecomments{Column 1 is the date of observation, column 2 the
observed band, column 3 the number of data points, column 4 time spans, column 5 the average $F$ value, column 6 the
critical $F$ value with 99 per cent confidence level, column 7 the
$F$ value of ANOVA, column 8 the critical $F$ value of ANOVA with 99
per cent confidence level, column 9 the variability status (V:
variable, PV: probable variable, N: non-variable), column 10-11 the
variability amplitude and daily average magnitudes respectively.}
\end{deluxetable}

\begin{deluxetable}{cccr}
\small
\tablecaption{Results of errors weighted Linear Regression Analysis. \label{tbl-4}} \tablewidth{0pt}
\tablehead{\colhead{Date(UT)} & \colhead{$N$} & \colhead{$r$} & \colhead{$P$}} \startdata
2016 Mar 29	&	20	&	0.55	&	0.01	\\
2016 Mar 30	&	50	&	0.42	&	0.002	\\
2016 Mar 31	&	51	&	0.36	&	0.01	\\
2016 Apr 23	&	50	&	0.24	&	0.09	\\
2016 Apr 24	&	90	&	0.31	&	0.003	\\
2016 Apr 25	&	95	&	0.35	&	$<0.0001$	\\
2016 Apr 26	&	96	&	0.39	&	$<0.0001$	\\
2016 May 09	&	31	&	0.52	&	0.003	\\
2016 May 10	&	18	&	0.63	&	0.005	\\
2016 Mar 29-2016 May 10	&	501	&	0.39	&	$<0.0001$	\\
2015 Apr 15	&	24	&	0.29	&	0.16	\\
2015 Apr 14	&	20	&	0.37	&	0.11	\\
2015 Apr 13	&	23	&	0.58	&	0.004	\\
2015 Apr 12	&	17	&	0.26	&	0.32	\\
2015 Apr 08	&	26	&	0.67	&	0.00017	\\
2015 May 17	&	17	&	0.44	&	0.08	\\
2015 May 19	&	12	&	0.72	&	0.009	\\
2015 Apr 08-2015 May 19	&	139	&	0.39	&	$<0.0001$	\\
2014 Apr 19	&	11	&	0.64	&	0.03	\\
2014 Apr 20	&	10	&	0.98	&	$<0.0001$	\\
2014 Apr 21	&	79	&	0.82	&	$<0.0001$	\\
2014 Apr 22	&	10	&	0.88	&	0.0008	\\
2014 Apr 23	&	12	&	0.6	&	0.04	\\
2014 Apr 24	&	14	&	0.59	&	0.025	\\
2014 Apr 25	&	17	&	0.35	&	0.16	\\
2014 Apr 26	&	12	&	0.84	&	0.0006	\\
2014 Apr 19-2014 Apr 26	&	165	&	0.61	&	$<0.0001$	\\
2013 Apr 01	&	15	&	0.3	&	0.28	\\
2013 Apr 04 	&	20	&	0.47	&	0.036	\\
2013 Apr 01-2013 Apr 04 	&	35	&	0.26	&	0.13	\\
2012 Feb 27-2012 Dec 21	&	47	&	0.5	&	0.0003	\\
2010 Apr 04 	&	30	&	0.57	&	0.001	\\
2010 Apr 04 -2010 May 03 	&	65	&	0.45	&	0.0002	\\
2009 Apr 15	&	32	&	0.66	&	$<0.0001$	\\
2009 Apr 14	&	24	&	0.68	&	0.0003	\\
2009 Mar 26-2009 Dec 14	&	84	&	0.87	&	$<0.0001$	\\
2005 Apr 06	&	15	&	0.82	&	0.0002	\\
2005 Apr 05-2005 Apr 06	&	19	&	0.89	&	$<0.0001$	\\
All data	&	1082	&	0.04	&	0.18	\\
\enddata
\tablecomments{$N$ is number of data; $r$ is the
coefficient of correlation; $P$ is the chance
probability.}
\end{deluxetable}

\end{document}